# Novel functions in silicon photonic chips incorporated with graphene oxide thin films

*David J. Moss*

[1]Optical Sciences Centre, Swinburne University of Technology, Hawthorn, VIC 3122, Australia.

*E-mail: *dmoss@swin.edu.au*

**Abstract**

On-chip integration of two-dimensional (2D) materials with unique structures and distinctive properties endow integrated devices with new functionalities and improved performance. With a high flexibility in modifying its properties and a strong compatibility with various integrated platforms, graphene oxide (GO) becomes an attractive 2D material for implementing functional hybrid integrated devices. Here, we demonstrate novel functionalities that go beyond the capabilities of conventional photonic integrated circuits, by harnessing the photo-thermal effects in 2D GO films integrated onto them. These include all-optical control and switching, optical power limiting, and non-reciprocal light transmission. The 2D layered GO films are on-chip integrated with precise control of their thicknesses and sizes. Benefitting from the broadband response of 2D GO films, all the three functionalities feature a very wide operational bandwidth. By fitting the experimental results with theory, we also extract the changes in material properties induced by the photo-thermal effects, which reveal interesting insights about 2D GO films. These results highlight the versatility of 2D GO films in implementing functional integrated photonic devices for a range of applications.

# Introduction

The advancement of integrated circuits, featured by compact footprint, low power consumption, high scalability, and the ability to significantly reduce cost by mass production, drives many technological breakthroughs in this information age[1,2]. Compared to their electronic counterparts, photonic integrated circuits (PICs) offer significantly greater processing bandwidth and robust immunity to electromagnetic interference, making them play crucial roles in a wide range of applications such as telecommunications, artificial intelligence, sensing, displays, and astronomy[3-6].

Silicon (Si) photonic platform that leverages the well-developed complementary metal-oxide-semiconductor (CMOS) fabrication technologies has been a leading platform for the state-of-the-art PICs[7,8]. Despite their wide deployment, Si PICs face intrinsic limitations arising from Si's material properties, presenting challenges in meeting the growing demands for device functionalities and performance. To address these challenges, on-chip integration of functional materials with superior properties has emerged as a promising solution[9-11].

Since the first experimental isolation of graphene in 2004[12], two-dimensional (2D) materials with atomically thin film structures and exceptional properties have garnered significant attention[13]. Among the various 2D materials, graphene oxide (GO) shows attractive advantages for implementing hybrid integrated photonic devices with high performance[14,15]. First, GO exhibits many excellent optical properties, such as ultrahigh optical nonlinearity, significant material anisotropy, and broadband response[16-18]. Second, there is a high flexibility in altering GO's properties through various reduction and doping methods, which substantially increases the variety of devices that can be developed[19,20]. Finally, GO has facile synthesis processes and transfer-free film coating with precise control over the thickness, showing a strong capability for large-scale on-chip integration[21,22].

Here, we propose and experimentally demonstrate novel functionalities for Si PICs enabled by on-chip integration of 2D GO films. By leveraging the unique property changes induced by photo-thermal effects in 2D GO films, three different functionalities are realized using GO-Si hybrid integrated waveguides, including all-optical control and switching, optical power limiting, and non-reciprocal light transmission. For all-optical control and switching, we first realize this functionality in non-resonant waveguides using continuous-wave (CW) light with low peak powers. For optical power limiting, we first demonstrate on-chip integration of a functional material to achieve power limiting for light propagation through integrated waveguides. For non-reciprocal light transmission, we first realize this functionality through on-chip integration of a 2D material, and our devices provide a considerably broader operation bandwidth that has not been achieved previously. Based on the experimental results, we further analyze the changes in the material properties induced by the photo-thermal effects, providing intriguing insights about 2D GO films. These results highlight the extensive opportunities arising from on-chip integration of 2D GO films for implementing functional integrated photonic devices.

## Results

### Device design and fabrication

**Fig. 1a** shows schematic of a Si waveguide integrated with a 2D GO film. The GO film is conformally coated onto the Si waveguide, which enables its strong interaction with the waveguide's evanescent field. Compared to integrated waveguides made from silicon nitride and doped silica[14], Si waveguides provide significantly increased optical mode overlaps with GO films. The enhanced mode overlap with GO allows for increased efficiency for the photo-thermal processes within it, consequently improving the device performance in the various applications we investigate in this study.

In the inset of **Fig. 1a**, we show the schematic of GO's atomic structure. As a common derivative of graphene, GO is comprised of carbon networks decorated with various oxygen-containing functional groups (OCFGs), such as hydroxyl, carboxyl, and carbonyl groups[15]. These OCFGs form bonds with certain carbon atoms in the carbon network through $sp^3$ hybridization, giving rise to a highly heterogeneous chemical structure[19]. Unlike graphene, which possesses a gapless Dirac cone[20], GO has an open bandgap (typically between 2 – 3.5 eV) that results from the presence of isolated $sp^2$ domains within the $sp^3$ C–O matrix[15]. The large bandgap of GO allows for a substantial reduction in its optical absorption, making it advantageous for optical applications that demand minimal loss[14,21]. In addition, the concentrations of $sp^2$ and $sp^3$ hybridizations in GO can be adjusted by reducing the OCFGs or doping treatments[22,23]. This provides a high level of flexibility in tailoring GO's properties, making it a versatile material platform for a variety of applications[15,21].

The transverse electric (TE) mode profile of the hybrid waveguide with a monolayer of GO is also shown in the inset. The GO layer is enclosed between two polymer layers that are incorporated during film coating based on self-assembly (See Methods), and such film structure is beneficial for enhancing the photo-thermal effects in the GO layer. The cross-section of the Si waveguide is 450 nm × 220 nm. We opt for TE-polarization here and in our subsequent experiments because it supports in-plane interaction between the waveguide's evanescent field and the 2D GO film. Due to the strong optical anisotropy in 2D materials, the in-plane interaction well surpasses the out-of-plane interaction in strength[17,24], thereby leading to an enhancement in the photo-thermal processes in GO and hence the device performance.

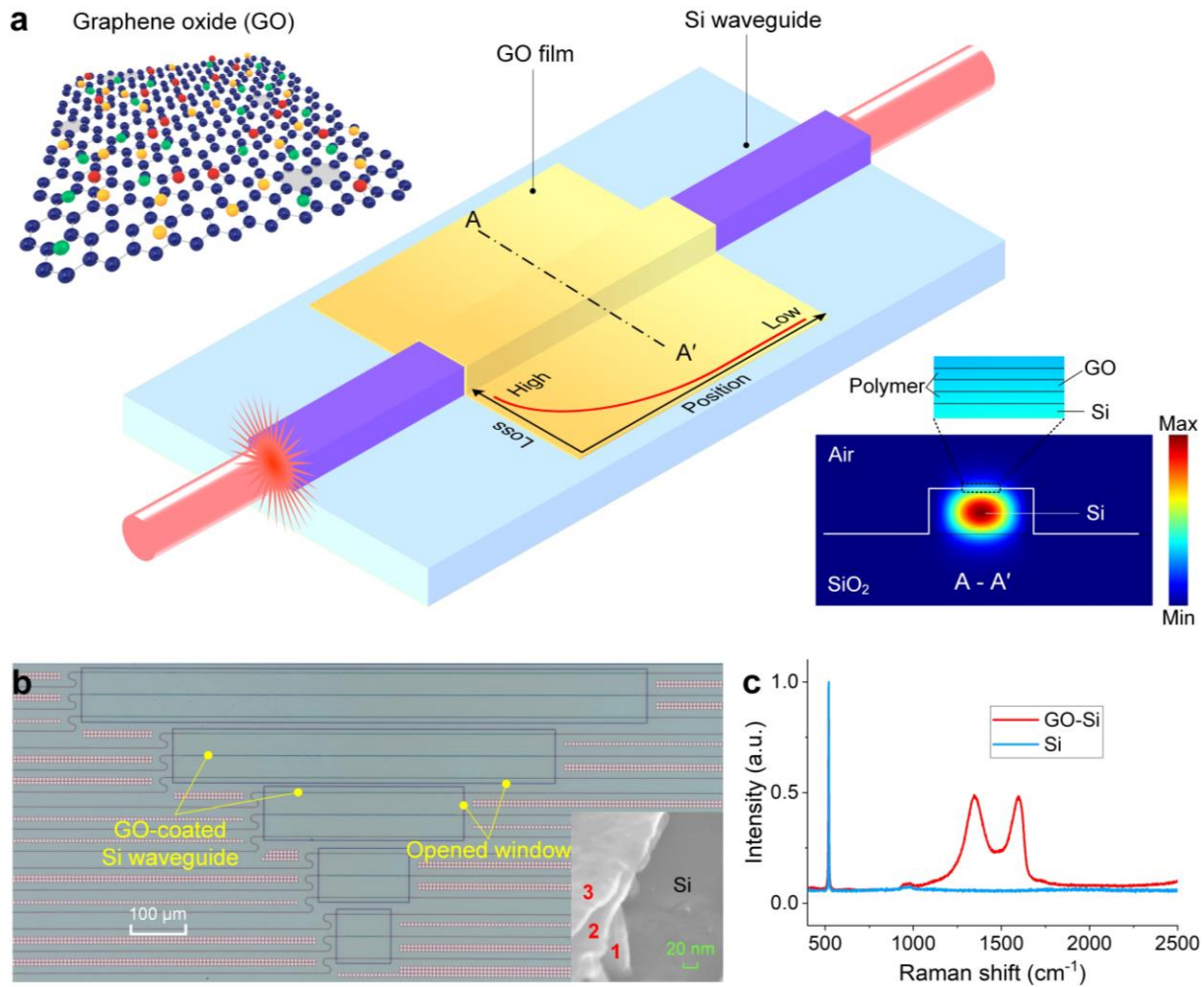

**Fig. 1 | Silicon (Si) waveguides integrated with 2D graphene oxide (GO) films. a**, Schematic illustration of a Si waveguide integrated with a 2D GO film. Insets show schematic of GO's atomic structure and transverse electric (TE) mode profile of the hybrid waveguide with a monolayer of GO. **b**, Microscopic image of fabricated Si chip coated with a monolayer of GO. Inset shows a scanning electron microscope (SEM) image of a 2D layered GO film coated on a Si substrate. The numbers 1−3 refer to the number of GO layers for that part of the image. **c**, Measured Raman spectra of the uncoated Si chip (Si) and the chip coated with a monolayer of GO (GO-Si).

**Fig. 1b** shows a microscopic image of a Si integrated chip coated with a monolayer of GO film. The uncoated Si waveguides were fabricated on a silicon-on-insulator (SOI) wafer using CMOS-compatible fabrication technologies (see Methods). To enable GO films coating onto the Si waveguides, windows were opened on the silica upper cladding of the integrated chip. The coating of the 2D GO film was achieved by using a solution-based self-assembly method that enables transfer-free and layer-by-layer film deposition (see Methods). A scanning electron microscope (SEM) image of a 2D layered GO film with up to three layers of GO on a Si substrate is shown in the inset. Our GO coating method allows for ultrathin GO film coating with a well-controlled thickness on a nanometer scale, offering both high repeatability and compatibility with various integrated material platforms[14,25,26]. The thickness of the as-prepared GO film, characterized via atomic force microscopy measurements, was ~2 nm / per layer. In addition, our approach can yield conformal film coating with direct contact and envelopment of GO films around the Si waveguides[27,28]. This results in efficient light-matter interaction and holds an advantage in contrast to film transfer methods typically employed for coating other 2D materials such as graphene and TMDCs[29]. The GO film's length can be controlled by adjusting the size of the opened window on the silica upper cladding, providing a way to optimize the device performance by changing the film length. In our fabricated chips, the length of all Si waveguides was ~3.0 mm, while the lengths of the opened windows varied from 0.1 mm to 2.2 mm. **Fig. 1c** shows the measured Raman spectra of a Si integrated chip before and after coating a GO film. The existence of the representative $D$ (1345 cm$^{-1}$) and $G$ (1590 cm$^{-1}$) peaks of GO in the Raman spectrum for the GO-coated chip confirms the successful on-chip integration of the GO film.

**Photo-thermal effects in 2D GO films**

In **Fig. 1a**, we illustrate that the loss in the GO-Si waveguide decreases along the direction of light propagation. Such decrease in loss occurs due to power-dependent photo-thermal processes within the GO film, which include a range of effects such as self-heating, thermal dissipation, and photo-thermal reduction[21]. These effects act together and mutually influence each other, resulting in alterations to GO's material properties such as refractive index and optical absorption. As mentioned previously, the reduction of the OCFGs in GO leads to changes in its properties. In the GO-Si waveguides, localized heating occurs due to optical absorption. Upon reaching a critical light power, the reduction of GO can be triggered when the thermal heating initiates a deoxygenation reaction in GO[30,31]. Such reaction leads to an increase in the optical absorption of the GO films[14] and hence the propagation loss of the hybrid waveguides. As a result, the propagation loss decreases in tandem with the attenuation of the light power along the GO-Si waveguides.

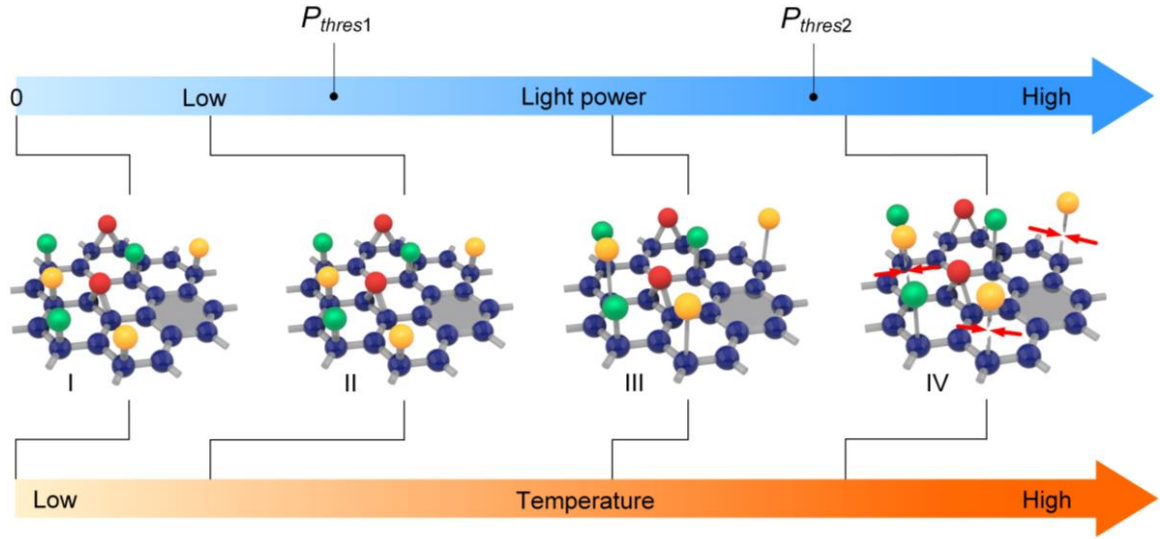

**Fig. 2 | Schematic illustration of changes in GO's atomic structure during photo-thermal processes in a GO film coated on a Si waveguide.** With rising light intensity in the GO-Si waveguide, the waveguide temperature increases, leading to photo-thermal effects that alter the atomic structure of GO and consequently modify its material properties. I – IV depict four stages, each corresponding to a different level of light power applied to the waveguide. $P_{thres1}$ and $P_{thres2}$ represent the power thresholds needed to initiate GO reduction and cause permanent changes in GO, respectively.

**Fig. 2** shows a schematic illustration of the changes in GO's atomic structure induced by the photo-thermal effects. Four stages, denoted as I – IV, correspond to different levels of light power applied to the GO-Si waveguide. Stage I represents the initial condition, with no incident light into the waveguide. At stage II, the light power in the waveguide is below $P_{thres1}$ – the threshold to initiate the reduction of GO, resulting in minimal and negligible changes in its properties. At stage III, GO reduction occurs when the light power in the waveguide exceeds $P_{thres1}$. An intriguing characteristic of this stage is the reversibility of GO reduction, meaning that when the light power is switched off, the waveguide can return to stage I with the same propagation loss. This reversibility is attributed to the fact that the reduction of GO induced by the photo-thermal effects is inherently unstable, and the reduced GO can readily revert to its initial state after cooling down. At stage IV, as the light power continues to rise, the temperature increase due to localized heating becomes sufficiently high to permanently break the chemical bonds between the OCFGs and the carbon network. This leads to a permanent change in the properties of the GO film, and the power threshold for initiating such a change is denoted as $P_{thres2}$. In this stage, the waveguide loss can decrease to some extent when the optical power is turned off, but it can no longer return to the loss level as observed in stage I.

In addition to the reversible photo-thermal reduction, another noteworthy characteristic of the photo-thermal effects in GO is their broad response bandwidth. This is enabled by the nearly flat absorption spectrum of GO that covers wavelengths from the visible to the infrared[14,25], as well as the heat-driven nature of these processes. The broadband response of these effects opens up the possibility of implementing various functional optical devices with broad operation bandwidths, as exemplified in our subsequent experimental demonstrations.

Previously[16,32], we observed relatively weak photo-thermal effects in GO films coated on silicon nitride and doped silica waveguides. In these waveguides used for nonlinear optical applications, the increased loss induced by photo-thermal reduction of GO deteriorated the device performance. In this study, from a different perspective, we harness the photo-thermal effects in GO films to realize several new functionalities for Si PICs. Si waveguides with much stronger mode overlaps with GO films coated on them facilitate significantly enhanced photo-thermal effects. In addition, we optimized the device design by incorporating a polymer cover layer over the coated GO films (See Methods), as opposed to the direct exposure of the GO films to air in our previous studies[16,17,32]. This can effectively prevent the escape of the OCFGs that are released during the photo-thermal processes and facilitate their chemical bonding back into the carbon network, thus expanding the power ranges for reversible GO reduction (see Note 7, Supplementary Information for detailed performance comparison).

## All-optical control and switching

Here we utilize the photo-thermal effects in GO to realize optical control and switching. **Fig. 3a** illustrates the operation principle, where a high-power pump light and a low-power probe light at different wavelengths were combined before injecting into a GO-Si waveguide with a monolayer of GO. After propagation through the hybrid waveguide, the pump light was filtered out, leaving only the probe light for characterization (see Note 1, Supplementary Information for detailed experimental setup).

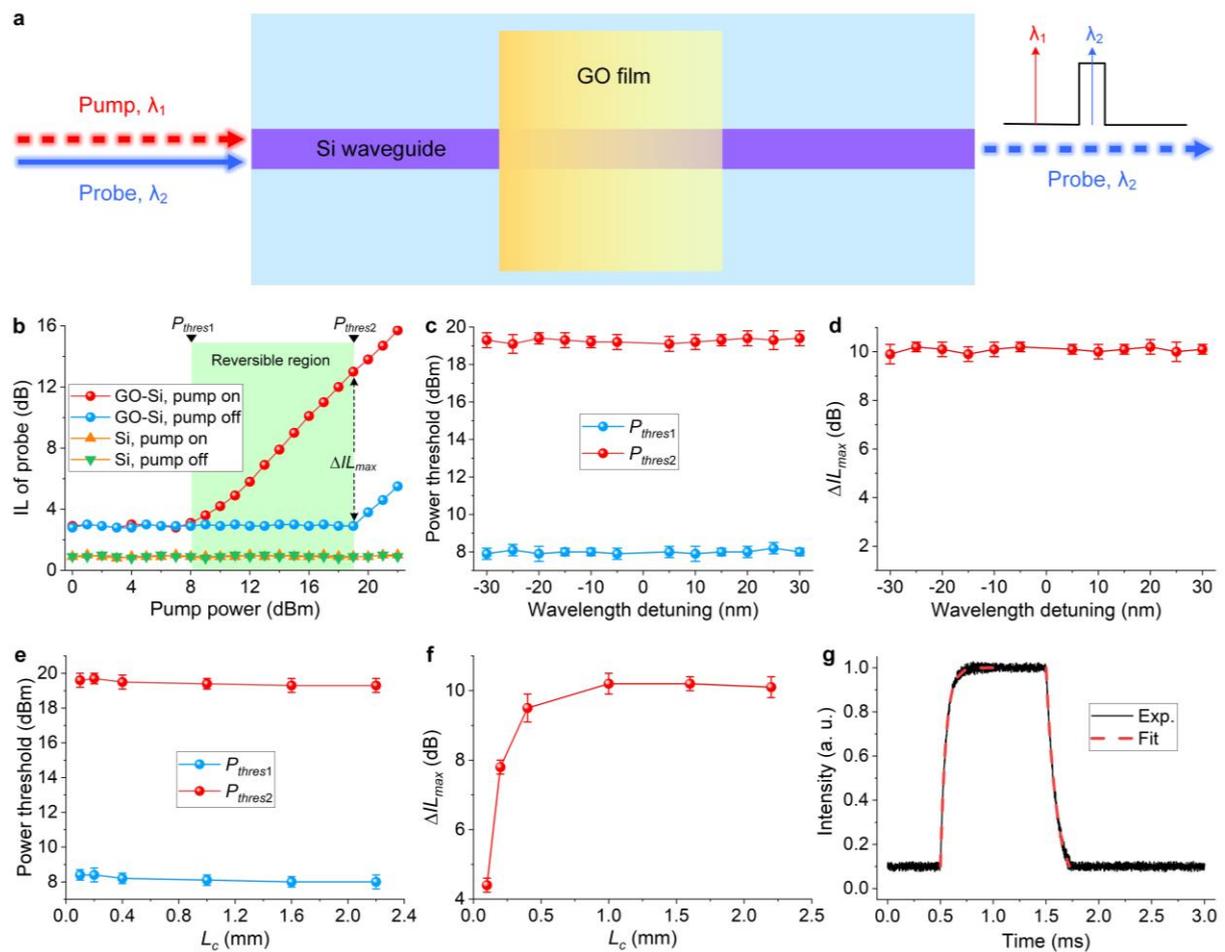

**Fig. 3 | Optical control and power switching in Si waveguides coated with a monolayer of GO. a,** Schematic of the principle. **b,** Measured insertion loss (IL) of probe light versus pump light power. The lower and upper limits of the pump power range for reversible photo-thermal changes are denoted as $P_{thres1}$ and $P_{thres2}$, respectively. At $P_{thres2}$, the variation in the IL between the pump being switched on and off is denoted as $\Delta IL_{max}$. The corresponding results for the uncoated Si waveguide are also shown for comparison. **c – d,** Measured $P_{thres1}$, $P_{thres2}$, and $\Delta IL_{max}$ versus wavelength detuning between the pump and probe. **e – f,** Measured $P_{thres1}$, $P_{thres2}$, and $\Delta IL_{max}$ versus GO coating length $L_c$. **g,** Measured and fit waveform of the probe when the pump is modulated by a 1-ms square electrical signal and then

amplified to ~19 dBm. In **b**, **c**, **d,** and **g**, $L_c$ = 1.0 mm. In **b**, **e**, **f**, and **g**, the wavelength detuning is ~30 nm.

**Fig. 3b** shows measured insertion loss (IL) of the probe light (excluding coupling loss between chip and fibres) versus power of the pump light (coupled into the waveguide). Both the pump and probe were CW light. As the pump power increased, the power of the probe light was maintained at a constant level of ~0 dBm. The 'pump on' results were recorded when the pump light was tuned on, and after the waveguide reached a steady thermal equilibrium state with no significant variations in the IL. Whereas the 'pump off' results were taken by initially activating the pump light and then deactivating it after the waveguide reached the thermal equilibrium state. For comparison, the corresponding results for the uncoated Si waveguide are also shown. For the GO-Si waveguide, the IL of the probe increased significantly when applying high pump powers. In contrast, the uncoated Si waveguide exhibited minimal variations in the IL. This confirms that the loss change was induced by the coated GO film. Note that the additional loss induced by two photon absorption (TPA) and free carrier absorption of Si is negligible (< 0.1 dB) within the power range we investigate here (see Note 2, Supplementary Information).

When the pump power was below ~8 dBm, the IL in **Fig. 3b** remained constant, indicating negligible variations in the optical absorption of the GO film. This reflects that the light power was insufficient to trigger any noticeable changes induced by the photo-thermal effects. When the pump power ranged between ~8 dBm and ~19 dBm, the IL of the probe increased as the pump power was raised. Upon deactivating the pump, the IL could return to the level observed when the pump power was below 8 dBm. These phenomena reflect that the changes induced by the photo-thermal effects in GO were reversible, with no permanent changes in the GO film after cooling it down. When the pump power exceeded ~20 dBm, the IL measured after turning off the pump no longer returned to the level at low powers but instead increased with the pump power, indicating that there were permanent changes in the GO film. The change in the intensity ratio of the *D* to *G* peaks observed in measured Raman spectra provides evidence that the GO at the start of the GO-coated segment experienced reduction (see Note 3, Supplementary Information). All of these observations align with the changes depicted in **Fig. 2**, providing evidence for the existence of photo-thermal effects in 2D GO films at high power levels.

In **Fig. 3b**, the lower and upper limits of the pump power range for the reversible photo-thermal changes are marked as $P_{thres1}$ and $P_{thres2}$, respectively, which align with those used in **Fig. 2**. At the pump power of $P_{thres2}$, the change in the IL between the pump being switched on and off are marked as $\Delta IL_{max}$. This represents the maximum variation in the IL within the power range associated with reversible photo-thermal changes. In **Figs. 3c-f**, we provide characterization for these three parameters by varying the wavelength tuning between the pump and probe and the GO coating length $L_c$. In each figure, the data points represent the average values obtained from the measured results of three duplicate devices and the error bars indicate the variations across different samples.

The results in **Figs. 3c** and **d** were measured using devices with the same $L_c$ of 0.4 mm. None of the three parameters display any significant variations within the investigated wavelength detuning range. This is attributed to the broadband response of the photo-thermal effects in the 2D GO films, as discussed earlier. Noted that in our experiments the range for wavelength detuning was constrained by the operation bandwidth of the optical amplifier, while the absorption response bandwidth for the GO films is, in fact, quite extensive, covering infrared wavelengths and even extending into the visible range[17]. The broadband response allows for optical control with the wavelengths of the pump and probe separated by several hundreds of nanometres or even higher, which is challenging for integrated photonic devices based on bulk materials[33].

**Figs. 3e** and **f** show the results measured at the same wavelength detuning of ~30 nm using devices with different $L_c$. As $L_c$ increases, both $P_{thres1}$ and $P_{thres2}$ remain nearly unchanged, with only minimal decreases due to the slightly reduced loss along the Si waveguides before reaching the start of the GO films (given that the windows of various lengths were opened at the center of the waveguides). This reflects the fact that the photo-thermal changes occurred primarily at the initial section of the GO film near the waveguide input and gradually diminished as the light power attenuated along the film. On the other hand, $\Delta IL_{max}$ initially increases with $L_c$, and then gradually levels off when $L_c$ becomes > 0.4 mm. This reflects that the variation in the IL caused by the photo-thermal effects accumulated over a specific GO film length between 0.4 mm and 1.0 mm (See Note 12 for detailed analysis). For GO films longer than this length, only the initial section of the films experienced substantial changes due to the photo-thermal effects. The remaining portion remained unaffected because the light power in it was insufficient to induce any significant photo-thermal changes.

We also demonstrated on–off power switching using the hybrid waveguide, where the pump light was modulated with a 1-ms square electrical signal and then amplified to ~19 dBm (see experimental setup in Note 4, Supplementary Information). We chose a GO film length of $L_c$ = 1.0 mm in order to minimize the loss of the unreduced portion and reduce the IL of the hybrid waveguide. As the pump light at ~1560 nm switched between the on and off states, the photo-thermal effects in GO resulted in corresponding changes of the probe light at ~1530 nm. The extinction ratio between the on and off states at the probe wavelength was ~10 dBm – consistent with the value of $\Delta IL_{max}$ at a pump power of ~19 dBm in **Fig. 3b**. In **Fig. 3g**, by fitting the temporal waveform of the probe light after propagation through the hybrid waveguide, we obtained the rising and falling time constants of ∼52 μs and ~98 μs, respectively (see Note 5, Supplementary Information). These values show agreement with the typical thermo-optic responses[31,34], providing further evidence that the changes were induced by the photo-thermal effects in GO.

Although all-optical switching with much faster time responses can be achieved based on saturable absorption (SA)[35], it usually necessitates optical pulses with high peak powers to trigger the SA, and in Si waveguides the high peak powers also induce significant TPA[5]. In contrast, our device in **Fig. 3a** operates effectively with CW light at considerably lower peak powers that do not cause any significant SA or TPA. This makes it attractive for applications that use CW light and do not demand ultrafast response times. In addition, our device is based on non-resonant waveguides that provide a very broad operation bandwidth. This distinguishes it from all-optical switching based on wavelength shifts of resonators[36], where the operation bandwidths are constrained by the resonance bandwidths of the resonators.

**Optical power limiting**

Here we employ the photo-thermal effects in GO to realize optical power limiting, which provides protection against potential damages induced by excessive light power[37]. As illustrated in **Fig. 4a**, the light with a higher power propagation through the GO-Si waveguide experiences a greater loss compared to the light with a lower power. Despite the difference between the input powers, the output powers remain nearly identical in both cases.

**Fig. 4b** compares the measured output power ($P_{out}$) versus input power ($P_{in}$) for a CW light propagation through the uncoated waveguide and the hybrid waveguide with a monolayer of GO (see experimental setup in Note 6, Supplementary Information). Note that the $P_{in}$ and $P_{out}$ mentioned here and in our subsequent discussions refers to the average power at the start and end of the 3-mm-long Si waveguides (excluding coupling loss between chip and fibres), respectively. As $P_{in}$ increased from ~0 to ~25 dBm, $P_{out}$ from the uncoated Si waveguide exhibited a linear increase. In contrast, $P_{out}$ from the hybrid waveguide showed a power limiting behaviour when $P_{in}$ exceeded ~8 dBm. Such behaviour was enabled by the photo-thermal effects in GO, which introduced additional propagation loss in the GO-coated segment. Due to the light-driven nature of the photo-thermal effects, the increase in the IL of the hybrid waveguide, once naturally reaching a stable thermal equilibrium state, was self-adjusting. It matched the rise in $P_{in}$, and consequently resulted in a limited $P_{out}$.

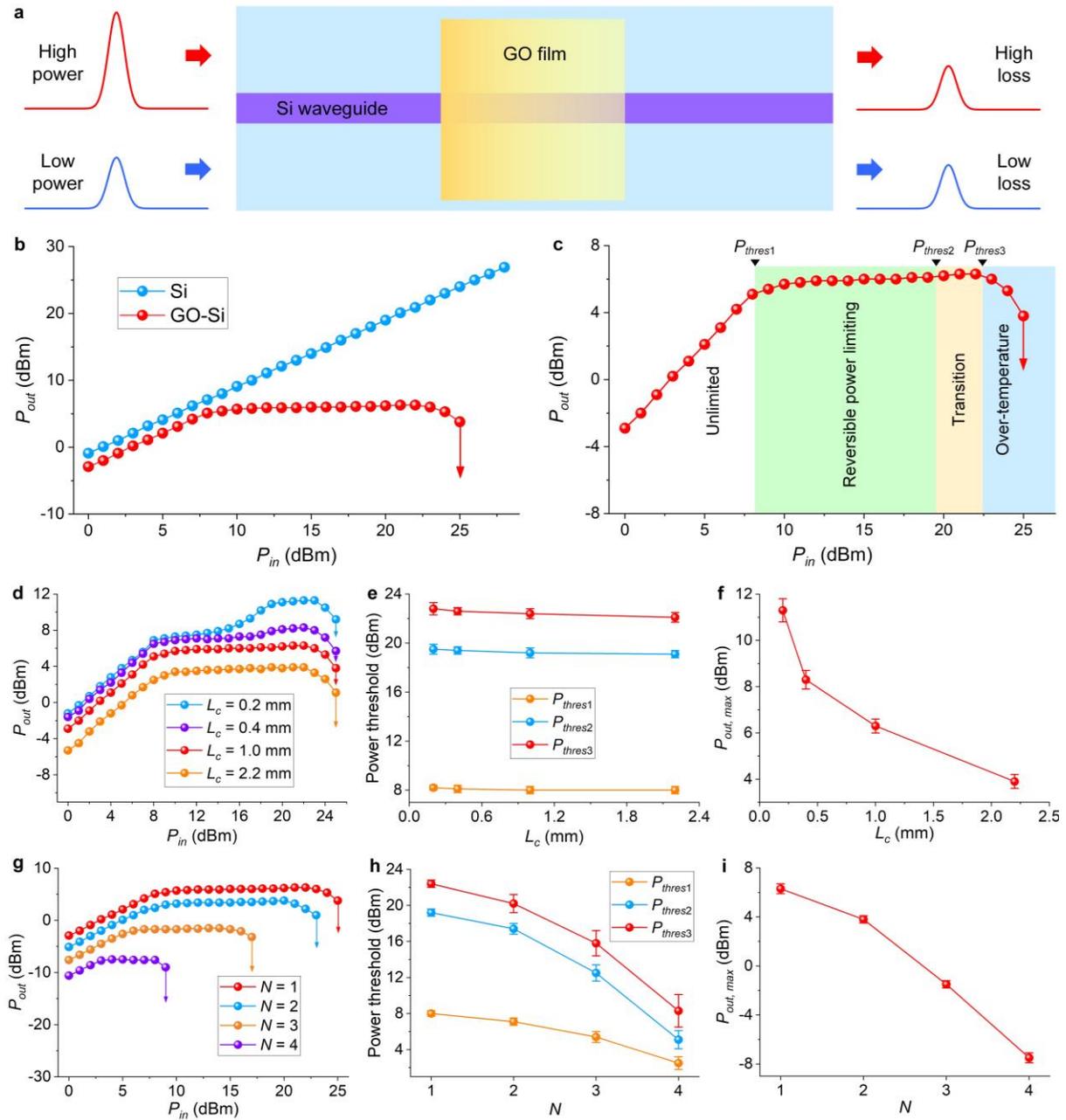

**Fig. 4 | Characterization of power limiting properties of 2D GO films coated on Si waveguides. a,** Schematic of the principle. **b,** Measured output power ($P_{out}$) versus input power ($P_{in}$) for a continuous-wave light propagation through an uncoated Si waveguide and a hybrid waveguide with a monolayer of GO. **c,** An enlarged view for the result of the hybrid waveguide in **b**, which is divided into four sections separated by $P_{thres1}$, $P_{thres2}$, and $P_{thres3}$. **d,** Measured $P_{out}$ versus $P_{in}$ for the waveguides with different GO coating lengths ($L_c$). **e,** Measured $P_{thres1}$, $P_{thres2}$, and $P_{thres3}$ versus $L_c$. **f,** Measured maximum output power ($P_{out, max}$) versus $L_c$. **g,** Measured $P_{out}$ versus $P_{in}$ for the waveguides with different numbers of GO layers ($N$). **h,** Measured $P_{thres1}$, $P_{thres2}$, and $P_{thres3}$ versus $N$. **f,** Measured $P_{out, max}$ versus $N$. In **b**, **c**, **d**, and **g**, the arrows indicate a rapid decrease in $P_{out}$ once reaching the corresponding input power thresholds. In **b** – **f**, $N$ = 1. In **b**, **c**, and **g** – **i**, $L_c$ = 1.0 mm.

In **Fig. 4c**, the measured curve for the hybrid waveguide in **Fig. 4b** is further divided into four sections, each corresponding to a different input power range. For $P_{in}$ below ~8 dBm, the power limiting phenomenon was not observed because the photo-thermal changes at such low powers are negligible. For $P_{in}$ in the range between ~8 dBm and ~19 dBm, where reversible photo-thermal changes occurred, $P_{out}$ increased very slightly when $P_{in}$ < 13 dBm and remained nearly constant when $P_{in}$ > 13 dBm. In addition, the power limiting behaviour was repeatable upon reinjection of the same light power. When $P_{in}$ ranged between ~20 dBm and ~22 dBm, despite the permanent changes in the GO film, the power limiting capability was still maintained. This is because the increase in loss arising from the permanent photo-thermal changes also adjusted itself to align with the increase in $P_{in}$. For $P_{in}$ exceeding ~23 dBm, a decline in the output power was observed due to the limited durability of the polymer layers that enclosed the GO layer (see Methods for GO film coating based on self-assembly), which were unable to endure the elevated temperatures at such high power levels. When $P_{in}$ reached ~25 dBm, the polymer layers suffered substantial thermal damage, resulting in a rapid drop in $P_{out}$. Nevertheless, even in this situation, the film still protected the output power from overloading.

The above characteristics allow the 2D GO films to function as fuses in integrated photonic devices, similar to their role in electronic circuits. It is worth noting that although a range of power limiting materials have been studied[38], the results here represent the first demonstration of on-chip integration of a functional material to achieve power limiting for light propagation through integrated waveguides. In addition, the broadband response of 2D GO films, along with their ease of removal (e.g., via plasma oxidation[17]) and subsequent recoating, enhances their practical utility over a wide wavelength range.

In **Figs. 4d–f**, we characterize the power limiting performance of the hybrid waveguides with different GO coating length $L_c$. For comparison, all the waveguides were coated with a monolayer of GO. As shown in **Fig. 4d**, for the waveguides with a $L_c$ smaller than 1.0 mm (*i.e.*, the same as that in **Fig. 4c**), the input power range exhibiting reversible power limiting behaviour is diminished. This is because at high powers the loss increase accumulated along the relatively small GO film length was insufficient to compensate for the rise in $P_{in}$. On the other hand, the input power range for power limiting does not further increase for the waveguide with a $L_c$ larger than 1.0 mm. This is due to the fact that, after propagation through the initial 1.0-mm-long GO film section, the light power becomes inadequate to induce substantial photo-thermal changes. In **Fig. 4e**, we provide characterization for $P_{thres1}$, $P_{thres2}$, and $P_{thres3}$, which are the input power thresholds dividing different sections in **Fig. 4c**. Similar to that in **Fig. 3e**, all of them exhibit minimal changes with increasing $L_c$, further confirming that the photo-thermal effects mainly took place at the initial portion of the GO film. As shown in **Fig. 4f**, the increase in $L_c$ leads to a decrease in the maximum output power from the hybrid waveguide. This is mainly due to the increased loss induced by a longer GO film length, and can be utilized to customize the output power according to various needs in practical applications.

**Figs. 4g–i** show characterization for the power limiting performance of hybrid waveguides coated with different numbers of GO layers ($N$ = 1– 4). For all the waveguides, $L_c$ = 1.0 mm. In **Fig. 4g**, the input power range with power limiting behaviour decreases for an increased $N$. In **Fig. 4h**, the three power thresholds also decrease as $N$ increases. These observations reflect the fact that the photo-thermal effects become more significant in thicker GO films, which can be attributed to the increased GO mode overlap and reduced thermal dissipation. In addition, the increase in impurities and defects within the GO layers, along with the unevenness and imperfect contact within the multi-layered film structure, result in reduced power endurance in thicker GO films. Similar to that in **Fig. 4f**, there is a decrease in the maximum output power as $N$ increases, mainly resulting from the increased loss for thicker GO films.

We compared the $P_{thres1}$, $P_{thres2}$, and $P_{thres3}$ for different waveguide configurations, including uncoated Si waveguides, Si waveguides coated with a polymer layer, Si waveguides coated with a polymer layer and a monolayer of GO, and Si waveguides coated with a monolayer of GO sandwiched between two polymer layers (see Note 7, Supplementary Information). The results show that the last configuration, which is also the one we employed to obtain the results in **Figs. 3–5**, provides the largest input power range for reversible photo-thermal changes. The performance for different light polarizations and Si waveguide geometries is also compared, as detailed in Notes 8 and 9, Supplementary Information. In practical applications, the values of $P_{thres1}$, $P_{thres2}$, and $P_{thres3}$ can be tailored by adjusting the input polarization and waveguide geometry to meet different requirements.

**Non-reciprocal light transmission**

In linear and time-invariant optical transmission media, as those commonly found in PICs, the Lorentz reciprocity theorem imposes constraints that prohibit the breaking of time-reversal symmetry and reciprocity[39]. Nevertheless, achieving non-reciprocal light transmission holds fundamental significance for PICs and forms the basis for functional devices like optical diodes[39-41] and isolators[42-44]. In previous reports, non-reciprocal light transmission was realized by introducing various nonlinear mechanisms into PICs, such as those based on magneto-optic materials[45,46], Brillouin scattering or optomechanically induced transparency[47,48], Kerr optical effect[43,46], acousto-optic modulation[42,44], Parity-time symmetric devices[49,50], and thermo-optic effects[40,41]. Although these methods have been successful in achieving high NTRs, challenges remain in achieving non-reciprocal light transmission over broad bandwidths. This is primarily because most of them rely on optical resonators with limited resonance bandwidths for enhancing the nonlinear response.

The photo-thermal effects in GO coated on integrated waveguides provide a new mechanism for realizing broadband nonreciprocal light transmission based on all-passive and non-resonant devices. As shown in **Fig. 5a**, a part of the GO film coated on a Si waveguide is permanently reduced. Compared to pristine GO, the permanently reduced GO (PR-GO) has significantly increased light absorption and does not show any obvious loss increase induced by the photo thermal effects[25,26]. For a high-power light traveling in the forward direction, it first goes through the PR-GO segment and subsequently enters the GO segment. After passing the PR-GO segment with substantial loss, the power entering the GO segment is insufficient to induce significant photo-thermal effects. In contrast, a light with the same power but propagating in the backward direction first encounters the GO segment, where it experiences additional loss due to the photo-thermal effects, before reaching the PR-GO segment with a consistent loss. Hence, it experiences higher loss compared to light traveling in the forward direction, leading to non-reciprocal light transmission. The broadband response of the photo-thermal effects in GO enables non-reciprocal light transmission over wide spectral ranges. Because the photo-thermal effects in GO are driven by light power in the hybrid waveguide, the nonreciprocal behaviour is influenced by the input power levels. This makes the hybrid waveguide cannot function as an optical isolator[39,51]. Nevertheless, the nonreciprocal transmission property can be effectively utilized in applications such as optical diodes, nonreciprocal switching, and signal processing[41,49].

In the experimental demonstration, we first measured a hybrid waveguide with a monolayer of GO. The PR-GO segment was fabricated via direct laser writing on the coated GO film with a length of $L_c$ = ~1.0 mm (see Methods). By adjusting the laser power and writing length, we achieved a consistent loss of ~10.6 dB for the PR-GO segment, which remained unaffected by a $P_{in}$ up to ~25 dBm. Such loss aligned with the value of $\Delta IL_{max}$ in **Fig. 3d**, which was aimed to improve the non-reciprocal transmission ratio (NTR) and minimize the extra loss caused by the PR-GO. **Fig. 5b** shows the measured waveguide IL as a function of wavelength for light propagating in both the forward and backward directions (see experimental setup in Note 10, Supplementary Information). The input power for light in both directions was kept the same as $P_{in}$ = ~19 dBm. Clearly, the light in the forward direction experienced a lower loss than that in the backward direction. The NTR, defined as the difference in the IL between light traveling in opposite directions, is further extracted and depicted in **Fig. 5c**. We obtained a flat NTR curve with uniform NTR values of ~10 dB across the entire C-band, which has not been achieved in previous reports (See detailed comparison in Note 11, Supplementary Information). In our demonstration, the wavelength tuning range was limited by the operation bandwidth of the optical amplifier. As previously mentioned, the photo-thermal effects in GO exhibit a much broader response bandwidth. This allows for a significantly wider bandwidth for non-reciprocal transmission than what was demonstrated in our experiments.

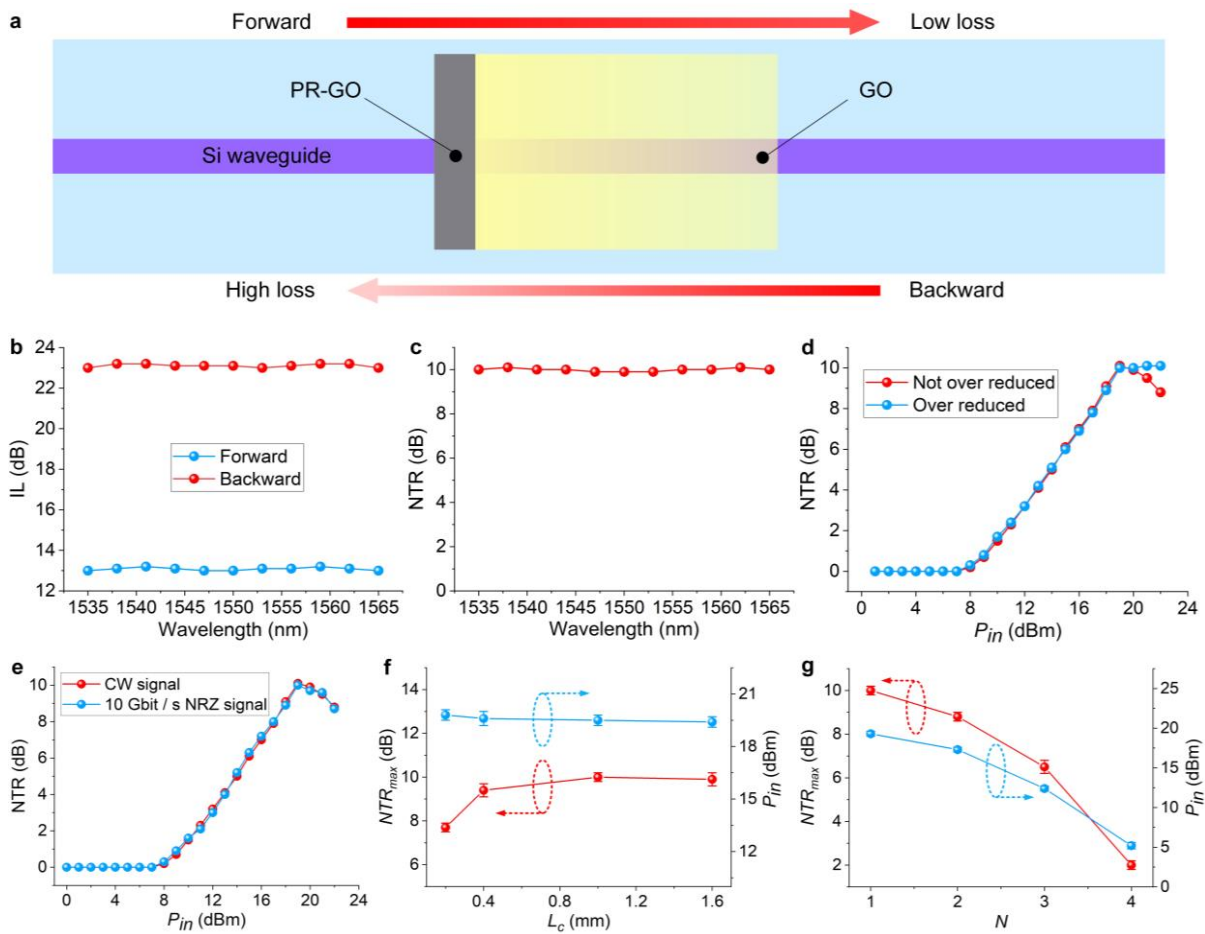

**Fig. 5 | Characterization of broadband non-reciprocal light transmission in GO-Si waveguides. a,** Schematic of the principle. A part of the GO film coated on a Si waveguide is permanently reduced (PR-GO) with a consistent loss unaffected by the light power. **b,** Measured insertion loss (IL) versus wavelength for continuous-wave (CW) light traveling in both the forward and backward directions. **c,**

Non-reciprocal transmission ratio (NTR) extracted from **b**. **d,** Measured NTR versus input power ($P_{in}$) when the PR-GO is over reduced or not. **e,** Measured NTR versus $P_{in}$ for both a CW light and a 10-Gbit/s non-return-to-zero (NRZ) signal. **f,** Measured maximum NTR ($NTR_{max}$) and corresponding $P_{in}$ versus GO coating length ($L_c$). **g,** Measured $NTR_{max}$ and corresponding $P_{in}$ versus GO layer number ($N$). In **b** – **e**, $L_c$ = 1.0 mm and $N$ = 1. In **f**, $N$ = 1. In **g**, $L_c$ = 1.0 mm.

In **Fig. 5d**, we show the NTR as a function of the input power $P_{in}$ for the hybrid waveguide measured in **Figs. 5b** and **c**. The corresponding result for another hybrid waveguide with over-reduced PR-GO is also shown for comparison. For the over-reduced PR-GO, we increased the length of the PR-GO segment to achieve a higher loss of ~15.0 dB, in contrast to ~10.6 dB for the PR-GO that was not over-reduced. For both cases, the NTR started to increase above 0 at $P_{in}$ = ~8 dBm and reached a maximum value of ~10.1 dB at $P_{in}$ = ~19 dBm. The power thresholds and the maximum NTR match with $P_{thres1}$, $P_{thres2}$, and $\Delta IL_{max}$ in **Fig. 3b**. For the waveguide with PR-GO that was not over-reduced, the NTR began to decrease as $P_{in}$ exceeded ~20 dBm. This occurs because the PR-GO segment was unable to entirely offset the surplus input power above $P_{thres2}$, allowing the forward-propagating light to still induce photo-thermal changes after passing through the PR-GO segment. Whereas for the waveguide with over-reduced PR-GO, despite having a higher IL, the maximum NTR could be sustained within a certain input power range where the excess input power can be adequately compensated. Note that when the input power of the backward-propagating light exceeded ~19 dBm, there was a permanent increase in the IL for both waveguides, but this did not impact the NTR values.

In **Fig. 5e**, we compare the non-reciprocal transmission performance for both a CW light and a 10-Gbit/s non-return-to-zero (NRZ) signal, using the device with PR-GO that was not over-reduced. Both the CW light and the NRZ signal exhibited similar NTRs at the same average power, reflecting that the photo-thermal effects are primarily influenced by the average power. Due to the substantial delay in the response of the photo-thermal effects compared to the rapid variations in the NRZ signal, the loss induced by the photo-thermal changes can be considered constant for the NRZ signal. This results in a similar NTR as observed with the CW light.

**Figs. 5f** and **g** show characterization for the non-reciprocal transmission performance of hybrid waveguides with different GO film lengths ($L_c$) and layer numbers ($N$), respectively. The change in the maximum NTR ($NTR_{max}$) with $L_c$ is consistent with the trend observed for the change of $\Delta IL_{max}$ in **Fig. 3f**. As $N$ increases, $NTR_{max}$ decreases, showing a trend similar to that observed for the maximum output power in **Fig. 4i**. In addition, the input powers to achieve $NTR_{max}$ values follow similar trends as those for $P_{thres2}$ in **Figs. 3** and **4**. These observations indicates that the three functionalities rely on the same fundamental mechanism of photo-thermal effects in 2D GO films.

**Theoretical analysis and discussion**

Based on the above experimental results, we theoretically model the photo-thermal changes in 2D GO films coated on integrated waveguides and analyze the film property changes with varying light power and temperature.

As the light power attenuates along the waveguide, the photo-thermal effects become weaker, leading to a smaller difference in properties between the photo-thermally reduced GO and the unreduced GO. This non-uniform behaviour represents an interesting characteristic for the photo-thermal changes in GO films coated on integrated waveguides. **Fig. 6a** shows a schematic illustration of a GO-Si waveguide with photo-thermal changes in GO induced by light propagation through. The GO film is divided into three segments with lengths of $L_{PR-GO}$, $L_{RR-GO}$, and $L_{NR-GO}$, which exhibit permanent reduction (PR), reversible reduction (RR), and no reduction (NR) behaviours, respectively. **Fig. 6b** shows the lengths of the three GO segments versus input power $P_{in}$, which was calculated based on the experimental results for the hybrid waveguide with $N = 1$ and $L_c = \sim 1.0$ mm in **Fig. 4** (see Note 12, Supplementary Information). As can be seen, for $P_{in} < \sim 8$ dBm, $L_{NR-GO}$ equals to $L_c$, indicating the presence of exclusively unreduced GO. Within the power range for reversible photo-thermal changes, $L_{RR\text{-}GO}$ increases with the input power, while $L_{NR\text{-}GO}$ shows an opposite trend, showing agreement with the observations in **Figs. 3b** and **4c**. In addition, $L_{PR\text{-}GO}$ remains at 0 until $P_{in}$ exceeds $P_{thres2}$, reflecting that the permanent changes only manifest within this specific power range. In a waveguide with permanently reduced GO, $L_{RR\text{-}GO}$ remains constant at ~0.43 mm. This constancy occurs because when the light power decreases to a level associated with reversible photo-thermal changes, the behaviour of GO closely resembles that observed within this power range, and $L_{RR\text{-}GO} = $ ~0.43 mm corresponds to $P_{in} = P_{thres2}$.

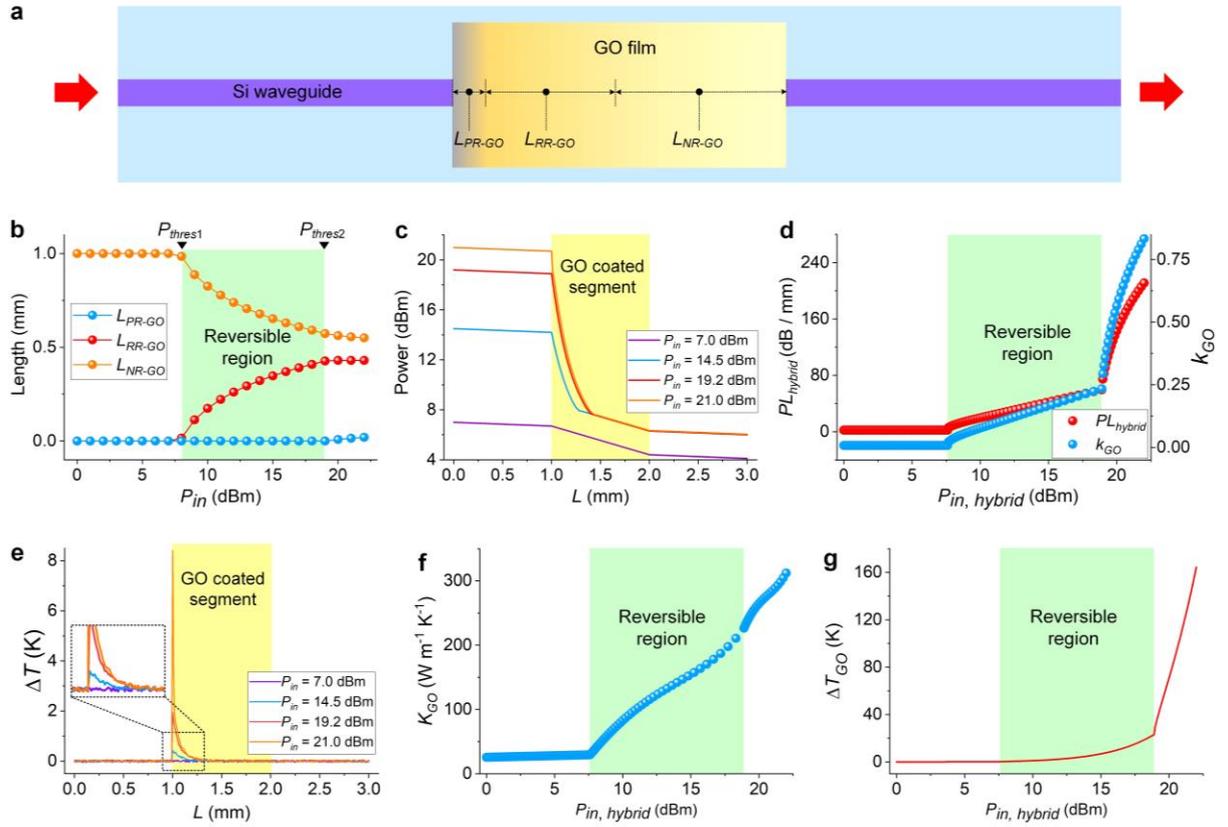

**Fig. 6 | Analysis for the photo-thermal changes in GO-Si waveguides. a,** Schematic of a GO-Si waveguide with photo-thermal changes, where the GO film is divided into three segments exhibiting permanent reduction (PR), reversible reduction (RR), and no reduction (NR) behaviours, with respective lengths denoted as $L_{PR\text{-}GO}$, $L_{RR\text{-}GO}$, and $L_{NR\text{-}GO}$. **b,** $L_{PR\text{-}GO}$, $L_{RR\text{-}GO}$, and $L_{NR\text{-}GO}$ versus input power ($P_{in}$). **c,** Evolution of different $P_{in}$ along the hybrid waveguide. **d,** Propagation loss of the hybrid waveguide ($PL_{hybrid}$) and extinction coefficient of GO ($k_{GO}$) versus light power injected into the hybrid waveguide cross-section ($P_{in,\,hybrid}$). **e,** Measured temperature distribution along the hybrid waveguide for different $P_{in}$. **f,** Thermal conductivity of GO ($K_{GO}$) versus $P_{in,\,hybrid}$ extracted from **e**. **g,** Simulated temperature variation at the GO film ($\Delta T_{GO}$) versus $P_{in,\,hybrid}$. In **b-f**, we show the results for the hybrid waveguide with a monolayer of GO ($N = 1$) at a coating length of $L_c = $ ~1.0 mm.

Based on the results in **Fig. 6b,** we calculated the evolution of light power along the hybrid waveguide (see Note 13, Supplementary Information). In **Fig. 6c**, we show the results for four different input power levels. At $P_{in}$ = ~7.0 dBm that is insufficient for any noticeable photo-thermal changes, there is only linear loss in both the uncoated and the GO-coated segments, which are ~0.3 dB/mm and ~2.3 dB/mm, respectively. For $P_{in}$ = ~14.5 dBm and ~19.2 dBm that fall within the range for reversible photo-thermal changes, the light power undergoes a substantial decrease in the RR-GO segment, followed by the same linear loss of ~2.3 dB/mm in the NR-GO segment. At $P_{in}$ = ~21.0 dBm, the light power first diminishes in the PR-GO segment to reach $P_{thres2}$ = ~19.2 dBm, and then experiences a loss similar to that observed for $P_{in}$ = ~19.2 dBm.

**Fig. 6d** shows the propagation loss of the hybrid waveguide ($PL_{hybrid}$) as a function of the light power injected into the hybrid waveguide cross-section ($P_{in,\ hybrid}$), which was calculated based on the results in **Fig. 6c** (see Note 14, Supplementary Information). Note that the $P_{in,\ hybrid}$ is directly related to $PL_{hybrid}$, and it is slightly different from $P_{in}$ in **Figs. 6b** and **c**. As $P_{in,\ hybrid}$ increases, $PL_{hybrid}$ first remains at a constant level, and then gradually increases in the power range for reversible photo-thermal changes. After that, there is a steep rise near the power threshold for permanent photo-thermal changes, followed by a subsequent gradual increase. In **Fig. 6d**, the extinction coefficient of the GO film ($k_{GO}$) is also plotted as a function of $P_{in,\ hybrid}$, which was extracted from the results for $PL_{hybrid}$ (see Note 14, Supplementary Information). The reversibly reduced GO exhibits higher light absorption compared to unreduced GO, and the permanently reduced GO shows even higher absorption. In a hybrid waveguide with significant photo-thermal effects, the reduced GO at the start of the waveguide absorbs a significant portion of light power, acting as a barrier to prevent reduction in the subsequent GO film. The degree of reduction in the reversibly reduced GO decreases as the light power attenuates along the film. This, in turn, amplifies the non-uniformity of the photo-thermal changes along the GO film.

**Fig. 6e** shows the temperature distributions along the hybrid waveguide for four different input powers $P_{in}$ that match those in **Fig. 6c**, which were measured by scanning a probe along the direction of the waveguide after it had achieved a stable thermal equilibrium state. The probe was positioned at a distance of ~2.5 μm from the bottom of the Si waveguide (*i.e.*, ~2.28 μm away from the GO film coated on the waveguide top surface). At $P_{in}$ = ~7.0 dBm, the temperature gradually decreases along the waveguide as the optical power attenuates, and there is no significant difference between the uncoated and GO-coated segments. As $P_{in}$ increases, the rise in the temperature at the start of the GO-coated segment (*i.e.*, $L$ = 1 mm) becomes more obvious, indicating the presence of more significant photo-thermal effects. The temperature experiences a rapid increase followed by a return to a typical level within the GO-coated segment, providing further evidence for the non-uniform photo-thermal changes in the GO film.

**Fig. 6f** shows the thermal conductivity of the 2D GO film ($K_{GO}$) as a function of $P_{in,\ hybrid}$, which was obtained by fitting the results in **Fig. 6e** with theoretical simulations (see Methods and Notes 15 and 16, Supplementary Information). In the fitting process, we also used the data from **Figs. 6c** and **d**. With an increase in $P_{in,\ hybrid}$, $K_{GO}$ also increases, and the increase becomes more obvious at higher $P_{in,\ hybrid}$ levels. These results, along with those in **Fig. 6d** for the extinction coefficient, reflect the transition from GO to reduced GO, which exhibits enhanced light absorption and increased thermal conductivity. At a high $P_{in,\ hybrid}$ of ~21.7 dB, $K_{GO}$ reaches ~312.4 Wm$^{-1}$K$^{-1}$. This value is much higher than the reported value for a 40-μm-thick reduced GO film[52], indicating a substantial increase in the thermal conductivity for reduced GO film in 2D form. We also note that this value is close to that reported for graphene on a dielectric substrate[53], which confirms the similarity in material properties between highly reduced GO and graphene.

The temperature distributions in **Fig. 6e** were measured using a probe positioned at a certain distance from the GO film. As a result, they cannot accurately reflect the temperature changes within the GO film during the photo-thermal processes. Based on the results in **Fig. 6f**, we performed simulations to estimate the temperature variations within the GO film (see Note 16, Supplementary Information). **Fig. 6g** shows the simulated temperature variation $\Delta T_{GO}$ as a function of $P_{in,\ hybrid}$. As expected, $\Delta T_{GO}$ increases with $P_{in,\ hybrid}$. The increase becomes more prominent within the power range for reversible reduction, and even more so within the power range for permanent reduction. At $P_{in,\ hybrid}$ = ~18.9 dB, $\Delta T_{GO}$ = ~23.0 K. We note that this value is lower compared to the reported values for much thicker GO films subjected to permanent reduction[52,54]. Such difference can be attributed to several factors, such as the much lower thicknesses of our films, the improved heat-trapping in the 2D GO layer enclosed by the polymer layers, and the existence of air voids in the layered film structure.

The various OCFGs in GO exhibit different bonding energies, resulting in the difference in their reduction temperatures[15]. Previous studies have found that hydroxyl (–OH) and carbonyl (C=O) groups possess much lower bonding energies compared to carboxylic (–COOH) and epoxy (–O–) groups[54]. Therefore, we infer that the reversible reduction of GO films at relatively low temperatures was primarily induced by changes in these two types of OCFGs. The difference between $P_{thres1}$ and $P_{thres2}$, which corresponds to the input power range for revisable photo-thermal changes in the GO films, is a crucial parameter in applications such as power limiting and non-reciprocal transmission. For instance, a larger difference between them enables a higher NTR. Further improving this parameter can be achieved by increasing the oxidation level of the unreduced GO, optimizing the film fabrication to better restrain the release of OCFGs, and minimizing the air voids and impurities within the multi-layered film structure. The thermal stability of the polymer layers (which enclose the GO layers) at high temperatures can also be improved by replacing the polymer used in our film coating (See

Methods) with polyimide, which has a notably high thermal stability[55]. These results have wide applications to GO based devices and other novel nonlinear materials. [56-82] Ultimately this could be useful for both classical and quantum microcomb based applications. [83-162]

## Conclusion

In summary, we harness the photo-thermal effects in 2D GO films integrated onto Si photonic waveguides to realize three new functionalities that are challenging for conventional PICs. These include all-optical control and switching, optical power limiting, and non-reciprocal light transmission. The 2D layered GO films are integrated onto Si photonic waveguides with precise control over their thicknesses, sizes, and placement. Owing to the broadband response of 2D GO films, all the three functionalities exhibit an exceptionally wide operational bandwidth. Based on the experimental results, we provide detailed analysis for the changes in material properties induced by the photo-thermal effects, providing insights into the characteristics of 2D GO films. These results provide a new route towards implementing high-performance functional integrated photonic devices though on-chip integration of 2D GO films.

## Methods

**Fabrication of Si waveguides.** The Si waveguides were fabricated on a SOI wafer with a 220-nm-thick top Si layer and a 2-μm-thick buried silicon dioxide ($SiO_2$) layer. The device layout was first defined on a negative photoresist using 248-nm deep ultraviolet photolithography and subsequently transferred to the top Si layer through photoresist development and inductively coupled plasma etching. Following this, a 1.5-μm-thick silica layer was deposited using plasma enhanced chemical vapor deposition to serve as the upper cladding layer. Finally, windows of different lengths were created down to the buried $SiO_2$ layer through the processes of photolithography and reactive ion etching. Lensed fibres were employed to butt couple light into and out of the fabricated Si waveguides with inverse-taper couplers at both ends. The coupling loss was ∼5 dB per facet.

**GO synthesis and film coating.** First, a GO solution composed of negatively charged GO flakes was prepared using a modified Hummers Method[56] that is facile and shows a high compatibility with CMOS fabrication. Vigorous sonication was employed to ensure that the dissolved GO flakes were in monolayer thickness and with lateral sizes < 100 nm. Second, the Si chip with a negatively charged surface was immersed in a 2.0% (w/v) polyelectrolyte polydiallyldimethylammonium chloride (PDDA) solution to obtain a polymer-coated integrated chip with a positively charged surface. Next, the polymer-coated Si chip was immersed in the prepared GO solution, where a GO monolayer was in-situ self-assembled onto the top surface through electrostatic forces. Finally, the Si chip with GO film on the top surface was re-immersed in the polymer solution to create a protective polymer top layer, shielding the GO film from direct exposure to air. The strong electrostatic forces enable conformal film coating with a high uniformity. Following the coating of each polymer or GO layer, the film surface was subjected to nitrogen gas blowing using an air gun. This was done to achieve a firm contact between the polymer and GO layers. By repeating the above steps, layer-by-layer coating of GO films can be realized, with high scalability and accurate control of the layer number or the film thickness[25,26]. After the GO film coating, the chip was dried in a drying oven at low temperature.

**Laser reduction of GO films.** The permanently reduced GO films in the non-reciprocal light transmission experiments were fabricated via direct laser writing using femtosecond optical pulses (~140-fs pulse width, ~800 nm wavelength) generated by an optical parametric oscillator. The laser writing system was modified on the basis of a Z-scan measurement system used in our previous studies[57,58] which has been used on a wide range of devices.[59-92] The generated laser beam was first expanded using a concave lens and two convex lenses, and then focused by an objective lens, resulting in a spot size of ~1.6 μm. The GO-Si waveguides were positioned perpendicular to the direction of the beam axis. The alignment of the light beam to the target position was achieved through a high-definition charge-coupled device (CCD) imaging system. To adjust the incident light power, a half-wave plate in conjunction with a linear polarizer was employed as a power attenuator. We did not fabricate permanently reduced GO induced by photo-thermal effects in the hybrid waveguides. This was because the difference between $P_{thres2}$ and $P_{thres3}$ was not sufficient to provide a loss of over 10 dB for the PR-GO segment. In addition, it proved challenging to precisely control the loss increase by raising the input power once the polymer layer sustained permanent thermal damage.

**Thermo-optic modeling and simulation.** The steady-state temperature distributions in the waveguide cross section were simulated using commercial finite-element multi-physics software to solve the heat equation described by the Fourier's law as

$$q = -K \cdot \nabla T \qquad (1)$$

where $q$ is the heat flux density, $K$ is the material's thermal conductivity, and $\nabla T$ is the temperature gradient. In our simulation, the heat power densities in different material regions were calculated individually, using the corresponding optical mode distribution and material property parameters (as detailed in Note 15, Supplementary Information). The values of $K$ for each of the material regions were also specified, with the exception of those that needed fitting. By fitting the experimental results with theoretical simulations, we obtained the fitted $K$ values of reduced GO at different input light powers. This was achieved by analyzing the measured temperature difference in the GO-coated segment (see details in Note 16, Supplementary Information). With the fitted $K$ values, the temperature variations within the GO film (as shown in **Fig. 6g**) were further calculated through thermal simulation using the same software. Due to the much lower mode overlap with GO on the sidewalls compared to that on the waveguide's top surface (see Note 17, Supplementary Information), our thermal simulations did not account for the variations in material properties arising from the anisotropy of 2D layered GO films, including parameters such as the extinction coefficient and thermal conductivity.

## Competing interests

The authors declare no competing interests.

## References

1    Bogaerts, W. *et al.* Programmable photonic circuits. *Nature* **586**, 207-216 (2020).
2    Liu, Y. *et al.* A photonic integrated circuit–based erbium-doped amplifier. *Science* **376**, 1309-1313 (2022).
3    Shen, Y. *et al.* Deep learning with coherent nanophotonic circuits. *Nat. Photonics* **11**, 441-446 (2017).
4    Chang, L., Liu, S. & Bowers, J. E. Integrated optical frequency comb technologies. *Nat. Photonics* **16**, 95-108 (2022).
5    Moss, D. J., Morandotti, R., Gaeta, A. L. & Lipson, M. New cmos-compatible platforms based on silicon nitride and hydex for nonlinear optics. *Nat. Photonics* **7**, 597-607 (2013).
6    Sun, Y. *et al.* Applications of optical microcombs. *Advances in Optics and Photonics* **15**, 86 (2023).
7    Rickman, A. The commercialization of silicon photonics. *Nat. Photonics* **8**, 579-582 (2014).
8    Reed, G. T., Mashanovich, G., Gardes, F. Y. & Thomson, D. J. Silicon optical modulators. *Nat. Photonics* **4**, 518-526 (2010).
9    Melikyan, A. *et al.* High-speed plasmonic phase modulators. *Nat. Photonics* **8**, 229-233 (2014).
10   Koos, C. *et al.* All-optical high-speed signal processing with silicon–organic hybrid slot waveguides. *Nat. Photonics* **3**, 216-219 (2009).
11   Sorianello, V. *et al.* Graphene–silicon phase modulators with gigahertz bandwidth. *Nat. Photonics* **12**, 40-44 (2017).
12   Novoselov, K. S. *et al.* Electric field effect in atomically thin carbon films. *Science* **306**, 666-669 (2004).
13   Koppens, F. H. *et al.* Photodetectors based on graphene, other two-dimensional materials and hybrid systems. *Nat. Nanotechnol.* **9**, 780-793 (2014).
14   Wu, J. *et al.* Graphene oxide for integrated photonics and flat optics. *Adv. Mater.* **33**, 2006415 (2021).
15   Wu, J., Lin, H., Moss, D. J., Loh, K. P. & Jia, B. Graphene oxide for photonics, electronics and optoelectronics. *Nature Reviews Chemistry* (2023).
16   Wu, J. *et al.* 2D layered graphene oxide films integrated with micro-ring resonators for enhanced nonlinear optics. *Small* **16**, 1906563 (2020).
17   Wu, J. *et al.* Graphene oxide waveguide and micro-ring resonator polarizers. *Laser Photonics Rev.* **13**, 1900056 (2019).

18. Shen, T. Z., Hong, S. H. & Song, J. K. Electro-optical switching of graphene oxide liquid crystals with an extremely large kerr coefficient. *Nat. Mater.* **13**, 394-399 (2014).
19. Loh, K. P., Bao, Q., Eda, G. & Chhowalla, M. Graphene oxide as a chemically tunable platform for optical applications. *Nat. Chem.* **2**, 1015-1024 (2010).
20. Bonaccorso, F., Sun, Z., Hasan, T. & Ferrari, A. C. Graphene photonics and optoelectronics. *Nat. Photonics* **4**, 611-622 (2010).
21. Zhang, Y. *et al.* Graphene oxide for nonlinear integrated photonics. *Laser Photonics Rev.*, 2200512 (2023).
22. Voiry, D. *et al.* High-quality graphene via microwave reduction of solution-exfoliated graphene oxide. *Science* **353**, 1413 (2016).
23. Guo, S., Garaj, S., Bianco, A. & Ménard-Moyon, C. Controlling covalent chemistry on graphene oxide. *Nature Reviews Physics* (2022).
24. Lin, H. *et al.* Chalcogenide glass-on-graphene photonics. *Nat. Photonics* **11**, 798-805 (2017).
25. Lin, H. *et al.* A 90-nm-thick graphene metamaterial for strong and extremely broadband absorption of unpolarized light. *Nat. Photonics* **13**, 270-276 (2019).
26. Lin, K. T., Lin, H., Yang, T. & Jia, B. Structured graphene metamaterial selective absorbers for high efficiency and omnidirectional solar thermal energy conversion. *Nat. Commun.* **11**, 1389 (2020).
27. Yang, Y. *et al.* Graphene metamaterial 3d conformal coating for enhanced light harvesting. *ACS Nano* (2022).
28. Zhang, Y. *et al.* Enhanced kerr nonlinearity and nonlinear figure of merit in silicon nanowires integrated with 2d graphene oxide films. *ACS Appl. Mater. Interfaces* **12**, 33094-33103 (2020).
29. Jia, L. *et al.* Fabrication technologies for the on-chip integration of 2d materials. *Small Methods*, 2101435 (2022).
30. Zhang, Y.-L. *et al.* Photoreduction of graphene oxides: Methods, properties, and applications. *Advanced Optical Materials* **2**, 10-28 (2014).
31. Chong, W. Y. *et al.* Photo-induced reduction of graphene oxide coating on optical waveguide and consequent optical intermodulation. *Sci. Rep.* **6**, 23813 (2016).
32. Qu, Y. *et al.* Enhanced four-wave mixing in silicon nitride waveguides integrated with 2d layered graphene oxide films. *Advanced Optical Materials* **8**, 2001048 (2020).
33. Ono, M. *et al.* Ultrafast and energy-efficient all-optical switching with graphene-loaded deep-subwavelength plasmonic waveguides. *Nat. Photonics* **14**, 37-43 (2019).
34. Li, Y. *et al.* Transforming heat transfer with thermal metamaterials and devices. *Nature Reviews Materials* **6**, 488-507 (2021).
35. Li, W. *et al.* Ultrafast all-optical graphene modulator. *Nano Lett.* **14**, 955-959 (2014).
36. Liu, F., Li, Q., Zhang, Z., Qiu, M. & Su, Y. Optically tunable delay line in silicon microring resonator based on thermal nonlinear effect. *IEEE J. Sel. Top. Quantum Electron.* **14**, 706-712 (2008).
37. Zhou, G. J. & Wong, W. Y. Organometallic acetylides of pt(ii), au(i) and hg(ii) as new generation optical power limiting materials. *Chem. Soc. Rev.* **40**, 2541-2566 (2011).
38. Pascal, S., David, S., Andraud, C. & Maury, O. Near-infrared dyes for two-photon absorption in the short-wavelength infrared: Strategies towards optical power limiting. *Chem. Soc. Rev.* **50**, 6613-6658 (2021).
39. Sounas, D. L. & Alù, A. Non-reciprocal photonics based on time modulation. *Nat. Photonics* **11**, 774-783 (2017).
40. Fan, L. *et al.* An all-silicon passive optical diode. *Science* **335**, 447-450 (2012).
41. Li, A. & Bogaerts, W. Reconfigurable nonlinear nonreciprocal transmission in a silicon photonic integrated circuit. *Optica* **7**, 7 (2020).
42. Tian, H. *et al.* Magnetic-free silicon nitride integrated optical isolator. *Nat. Photonics* **15**, 828-836 (2021).
43. White, A. D. *et al.* Integrated passive nonlinear optical isolators. *Nat. Photonics* (2022).
44. Kittlaus, E. A. *et al.* Electrically driven acousto-optics and broadband non-reciprocity in silicon photonics. *Nat. Photonics* **15**, 43-52 (2020).
45. Bi, L. *et al.* On-chip optical isolation in monolithically integrated non-reciprocal optical resonators. *Nat. Photonics* **5**, 758-762 (2011).
46. Yang, K. Y. *et al.* Inverse-designed non-reciprocal pulse router for chip-based lidar. *Nat. Photonics* **14**, 369-374 (2020).
47. Kim, J., Kuzyk, M. C., Han, K., Wang, H. & Bahl, G. Non-reciprocal brillouin scattering induced transparency. *Nat. Phys.* **11**, 275-280 (2015).
48. Shen, Z. *et al.* Experimental realization of optomechanically induced non-reciprocity. *Nat. Photonics* **10**, 657-661 (2016).
49. Feng, L. *et al.* Nonreciprocal light propagation in a silicon photonic circuit. *Science* **333**, 729-733 (2011).
50. Peng, B. *et al.* Parity–time-symmetric whispering-gallery microcavities. *Nat. Phys.* **10**, 394-398 (2014).
51. Jalas, D. *et al.* What is — and what is not — an optical isolator. *Nat. Photonics* **7**, 579-582 (2013).
52. Renteria, J. D. *et al.* Strongly anisotropic thermal conductivity of free-standing reduced graphene oxide films annealed at high temperature. *Adv. Funct. Mater.* **25**, 4664-4672 (2015).
53. Seol, J. H. *et al.* Two-dimensional phonon transport in supported graphene. *Science* **328**, 213-216 (2010).

54  Sengupta, I., Chakraborty, S., Talukdar, M., Pal, S. K. & Chakraborty, S. Thermal reduction of graphene oxide: How temperature influences purity. *J. Mater. Res.* **33**, 4113-4122 (2018).

55  Zhang, L. *et al.* Free-standing functional graphene reinforced carbon films with excellent mechanical properties and superhydrophobic characteristic. *Composites Part A: Applied Science and Manufacturing* **74**, 96-106 (2015).

56  Yang, Y. Y. *et al.* Graphene-based multilayered metamaterials with phototunable architecture for on-chip photonic devices. *ACS Photonics* **6**, 1033-1040 (2019).

57  Zheng, X., Jia, B., Chen, X. & Gu, M. In situ third-order non-linear responses during laser reduction of graphene oxide thin films towards on-chip non-linear photonic devices. *Adv. Mater.* **26**, 2699-2703 (2014).

58  Jia, L. *et al.* Highly nonlinear BiOBr nanoflakes for hybrid integrated photonics. *APL Photonics* **4**, 090802 (2019).

59. Yuning Zhang, Jiayang Wu, Yang Qu, Yunyi Yang, Linnan Jia, Baohua Jia, and David J. Moss, "Enhanced supercontinuum generated in SiN waveguides coated with GO films", Advanced Materials Technologies **8** (1) 2201796 (2023).

60. Yuning Zhang, Jiayang Wu, Linnan Jia, Yang Qu, Baohua Jia, and David J. Moss, "Graphene oxide for nonlinear integrated photonics", Laser and Photonics Reviews **17** 2200512 (2023). DOI:10.1002/lpor.202200512.

61. Jiayang Wu, H. Lin, D. J. Moss, T.K. Loh, Baohua Jia, "Graphene oxide: new opportunities for electronics, photonics, and optoelectronics", Nature Reviews Chemistry **7** (3) 162–183 (2023). DOI:10.1038/s41570-022-00458-7.

62. Yang Qu, Jiayang Wu, Yuning Zhang, Yunyi Yang, Linnan Jia, Baohua Jia, and David J. Moss, "Photo thermal tuning in GO-coated integrated waveguides", Micromachines **13** 1194 (2022). doi.org/10.3390/mi13081194

63. Yuning Zhang, Jiayang Wu, Yunyi Yang, Yang Qu, Houssein El Dirani, Romain Crochemore, Corrado Sciancalepore, Pierre Demongodin, Christian Grillet, Christelle Monat, Baohua Jia, and David J. Moss, "Enhanced self-phase modulation in silicon nitride waveguides integrated with 2D graphene oxide films", IEEE Journal of Selected Topics in Quantum Electronics **29** (1) 5100413 (2023). DOI: 10.1109/JSTQE.2022.3177385

64. Yuning Zhang, Jiayang Wu, Yunyi Yang, Yang Qu, Linnan Jia, Baohua Jia, and David J. Moss, "Enhanced spectral broadening of femtosecond optical pulses in silicon nanowires integrated with 2D graphene oxide films", Micromachines **13** 756 (2022). DOI:10.3390/mi13050756.

65. Linnan Jia, Jiayang Wu, Yuning Zhang, Yang Qu, Baohua Jia, Zhigang Chen, and David J. Moss, "Fabrication Technologies for the On-Chip Integration of 2D Materials", Small: Methods **6**, 2101435 (2022). DOI:10.1002/smtd.202101435.

66. Yuning Zhang, Jiayang Wu, Yang Qu, Linnan Jia, Baohua Jia, and David J. Moss, "Design and optimization of four-wave mixing in microring resonators integrated with 2D graphene oxide films", Journal of Lightwave Technology **39** (20) 6553-6562 (2021). DOI:10.1109/JLT.2021.3101292. Print ISSN: 0733-8724, Online ISSN: 1558-2213 (2021).

67. Yuning Zhang, Jiayang Wu, Yang Qu, Linnan Jia, Baohua Jia, and David J. Moss, "Optimizing the Kerr nonlinear optical performance of silicon waveguides integrated with 2D graphene oxide films", Journal of Lightwave Technology 39 (14) 4671-4683 (2021). DOI: 10.1109/JLT.2021.3069733.

68. Yang Qu, Jiayang Wu, Yuning Zhang, Yao Liang, Baohua Jia, and David J. Moss, "Analysis of four-wave mixing in silicon nitride waveguides integrated with 2D layered graphene oxide films", Journal of Lightwave Technology **39** (9) 2902-2910 (2021). DOI: 10.1109/JLT.2021.3059721.

69. Jiayang Wu, Linnan Jia, Yuning Zhang, Yang Qu, Baohua Jia, and David J. Moss," Graphene oxide: versatile films for flat optics to nonlinear photonic chips", Advanced Materials **33** (3) 2006415, pp.1-29 (2021). DOI:10.1002/adma.202006415.

70. Y. Qu, J. Wu, Y. Zhang, L. Jia, Y. Yang, X. Xu, S. T. Chu, B. E. Little, R. Morandotti, B. Jia, and D. J. Moss, "Graphene oxide for enhanced optical nonlinear performance in CMOS compatible integrated devices", Paper No. 11688-30, PW21O-OE109-36, 2D Photonic Materials and Devices IV, SPIE Photonics West, San Francisco CA March 6-11 (2021). doi.org/10.1117/12.2583978

71. Yang Qu, Jiayang Wu, Yunyi Yang, Yuning Zhang, Yao Liang, Houssein El Dirani, Romain Crochemore, Pierre Demongodin, Corrado Sciancalepore, Christian Grillet, Christelle Monat, Baohua Jia, and David J. Moss, "Enhanced nonlinear four-wave mixing in silicon nitride waveguides integrated with 2D layered graphene oxide films", Advanced Optical Materials vol. **8** (21) 2001048 (2020). DOI: 10.1002/adom.202001048. arXiv:2006.14944.

72. Yuning Zhang, Yang Qu, Jiayang Wu, Linnan Jia, Yunyi Yang, Xingyuan Xu, Baohua Jia, and David J. Moss, "Enhanced Kerr nonlinearity and nonlinear figure of merit in silicon nanowires integrated with 2D graphene oxide films", ACS Applied Materials and Interfaces vol. **12** (29) 33094–33103 June 29 (2020). DOI:10.1021/acsami.0c07852

73. Jiayang Wu, Yunyi Yang, Yang Qu, Yuning Zhang, Linnan Jia, Xingyuan Xu, Sai T. Chu, Brent E. Little, Roberto Morandotti, Baohua Jia,* and David J. Moss*, "Enhanced nonlinear four-wave mixing in microring resonators integrated with layered graphene oxide films", Small vol. **16** (16) 1906563 April 23 (2020). DOI: 10.1002/smll.201906563

74. Jiayang Wu, Yunyi Yang, Yang Qu, Xingyuan Xu, Yao Liang, Sai T. Chu, Brent E. Little, Roberto Morandotti, Baohua Jia, and David J. Moss, "Graphene oxide waveguide polarizers and polarization selective micro-ring resonators", Paper 11282-29, SPIE Photonics West, San Francisco, CA, 4 - 7 February (2020). doi: 10.1117/12.2544584

75. Jiayang Wu, Yunyi Yang, Yang Qu, Xingyuan Xu, Yao Liang, Sai T. Chu, Brent E. Little, Roberto Morandotti, Baohua Jia, and David J. Moss, "Graphene oxide waveguide polarizers and polarization selective micro-ring resonators", Laser and Photonics Reviews vol. **13** (9) 1900056 (2019). DOI:10.1002/lpor.201900056.
76. Yunyi Yang, Jiayang Wu, Xingyuan Xu, Sai T. Chu, Brent E. Little, Roberto Morandotti, Baohua Jia, and David J. Moss, "Enhanced four-wave mixing in graphene oxide coated waveguides", Applied Physics Letters Photonics vol. **3** 120803 (2018). doi: 10.1063/1.5045509.
77. Linnan Jia, Yang Qu, Jiayang Wu, Yuning Zhang, Yunyi Yang, Baohua Jia, and David J. Moss, "Third-order optical nonlinearities of 2D materials at telecommunications wavelengths", ***Micromachines*** (MDPI), **14**, 307 (2023). https://doi.org/10.3390/mi14020307.
78. Linnan Jia, Dandan Cui, Jiayang Wu, Haifeng Feng, Tieshan Yang, Yunyi Yang, Yi Du, Weichang Hao, Baohua Jia, David J. Moss, "BiOBr nanoflakes with strong nonlinear optical properties towards hybrid integrated photonic devices", Applied Physics Letters Photonics vol. **4** 090802 (2019). DOI: 10.1063/1.5116621
79. Linnan Jia, Jiayang Wu, Yunyi Yang, Yi Du, Baohua Jia, David J. Moss, "Large Third-Order Optical Kerr Nonlinearity in Nanometer-Thick PdSe2 2D Dichalcogenide Films: Implications for Nonlinear Photonic Devices", ACS Applied Nano Materials vol. **3** (7) 6876–6883 (2020). DOI:10.1021/acsanm.0c01239.
80. M Ferrera et al., "On-Chip ultra-fast 1st and 2nd order CMOS compatible all-optical integration", Optics Express vol. **19** (23), 23153-23161 (2011).
81. C Monat, C Grillet, B Corcoran, DJ Moss, BJ Eggleton, TP White, …, et al., "Investigation of phase matching for third-harmonic generation in silicon slow light photonic crystal waveguides using Fourier optics", Optics Express **18** (7), 6831-6840 (2010).
82. L Carletti, P Ma, Y Yu, B Luther-Davies, D Hudson, C Monat, …. , et al., "Nonlinear optical response of low loss silicon germanium waveguides in the mid-infrared", Optics Express **23** (7), 8261-8271 (2015).
83. Bao, C., et al., Direct soliton generation in microresonators, Opt. Lett, **42**, 2519 (2017).
84. M.Ferrera et al., "CMOS compatible integrated all-optical RF spectrum analyzer", Optics Express, vol. 22, no. 18, 21488 - 21498 (2014).
85. M. Kues, et al., "Passively modelocked laser with an ultra-narrow spectral width", Nature Photonics, vol. 11, no. 3, pp. 159, 2017.
86. L. Razzari, et al., "CMOS-compatible integrated optical hyper-parametric oscillator," Nature Photonics, vol. 4, no. 1, pp. 41-45, 2010.
87. M. Ferrera, et al., "Low-power continuous-wave nonlinear optics in doped silica glass integrated waveguide structures," Nature Photonics, vol. 2, no. 12, pp. 737-740, 2008.
88. M.Ferrera et al."On-Chip ultra-fast 1st and 2nd order CMOS compatible all-optical integration", Opt. Express, vol. 19, (23)pp. 23153-23161 (2011).
89. D. Duchesne, M. Peccianti, M. R. E. Lamont, et al., "Supercontinuum generation in a high index doped silica glass spiral waveguide," Optics Express, vol. 18, no, 2, pp. 923-930, 2010.
90. H Bao, L Olivieri, M Rowley, ST Chu, BE Little, R Morandotti, DJ Moss, … "Turing patterns in a fiber laser with a nested microresonator: Robust and controllable microcomb generation", Physical Review Research **2** (2), 023395 (2020).
91. M. Ferrera, et al., "On-chip CMOS-compatible all-optical integrator", Nature Communications, vol. 1, Article 29, 2010.
92. A. Pasquazi, et al., "All-optical wavelength conversion in an integrated ring resonator," Optics Express, vol. 18, no. 4, pp. 3858-3863, 2010.
93. A.Pasquazi, Y. Park, J. Azana, et al., "Efficient wavelength conversion and net parametric gain via Four Wave Mixing in a high index doped silica waveguide," Optics Express, vol. 18, no. 8, pp. 7634-7641, 2010.
94. M. Peccianti, M. Ferrera, L. Razzari, et al., "Subpicosecond optical pulse compression via an integrated nonlinear chirper," Optics Express, vol. 18, no. 8, pp. 7625-7633, 2010.
95. Little, B. E. et al., "Very high-order microring resonator filters for WDM applications", IEEE Photonics Technol. Lett. **16**, 2263–2265 (2004).
96. M. Ferrera et al., "Low Power CW Parametric Mixing in a Low Dispersion High Index Doped Silica Glass Micro-Ring Resonator with Q-factor > 1 Million", Optics Express, vol.17, no. 16, pp. 14098–14103 (2009).
97. M. Peccianti, et al., "Demonstration of an ultrafast nonlinear microcavity modelocked laser", Nature Communications, vol. 3, pp. 765, 2012.
98. A.Pasquazi, et al., "Self-locked optical parametric oscillation in a CMOS compatible microring resonator: a route to robust optical frequency comb generation on a chip," Optics Express, vol. 21, no. 11, pp. 13333-13341, 2013.
99. A.Pasquazi, et al., "Stable, dual mode, high repetition rate mode-locked laser based on a microring resonator," Optics Express, vol. 20, no. 24, pp. 27355-27362, 2012.

100. Pasquazi, A. et al. Micro-combs: a novel generation of optical sources. Physics Reports **729**, 1-81 (2018).
101. Moss, D. J. et al., "New CMOS-compatible platforms based on silicon nitride and Hydex for nonlinear optics", Nature photonics **7**, 597 (2013).
102. H. Bao, et al., Laser cavity-soliton microcombs, Nature Photonics, vol. 13, no. 6, pp. 384-389, Jun. 2019.
103. Antonio Cutrona, Maxwell Rowley, Debayan Das, Luana Olivieri, Luke Peters, Sai T. Chu, Brent L. Little, Roberto Morandotti, David J. Moss, Juan Sebastian Totero Gongora, Marco Peccianti, Alessia Pasquazi, "High Conversion Efficiency in Laser Cavity-Soliton Microcombs", Optics Express Vol. 30, Issue 22, pp. 39816-39825 (2022). https://doi.org/10.1364/OE.470376.
104. M.Rowley, P.Hanzard, A.Cutrona, H.Bao, S.Chu, B.Little, R.Morandotti, D. J. Moss, G. Oppo, J. Gongora, M. Peccianti and A. Pasquazi, "Self-emergence of robust solitons in a micro-cavity", Nature **608** (7922) 303–309 (2022).
105. A. Cutrona, M. Rowley, A. Bendahmane, V. Cecconi,L. Peters, L. Olivieri, B. E. Little, S. T. Chu, S. Stivala, R. Morandotti, D. J. Moss, J. S. Totero-Gongora, M. Peccianti, A. Pasquazi, "Nonlocal bonding of a soliton and a blue-detuned state in a microcomb laser", **Nature Communications Physics 6** (2023).
106. A. Cutrona, M. Rowley, A. Bendahmane, V. Cecconi,L. Peters, L. Olivieri, B. E. Little, S. T. Chu, S. Stivala, R. Morandotti, D. J. Moss, J. S. Totero-Gongora, M. Peccianti, A. Pasquazi, "Stability Properties of Laser Cavity-Solitons for Metrological Applications", **Applied Physics Letters 122** (12) 121104 (2023); doi: 10.1063/5.0134147.X. Xu, J. Wu, M. Shoeiby, T. G. Nguyen, S. T. Chu, B. E. Little, R. Morandotti, A. Mitchell, and D. J. Moss, "Reconfigurable broadband microwave photonic intensity differentiator based on an integrated optical frequency comb source," APL Photonics, vol. 2, no. 9, 096104, Sep. 2017.
107. Xu, X., et al., Photonic microwave true time delays for phased array antennas using a 49 GHz FSR integrated micro-comb source, *Photonics Research*, **6**, B30-B36 (2018).
108. X. Xu, M. Tan, J. Wu, R. Morandotti, A. Mitchell, and D. J. Moss, "Microcomb-based photonic RF signal processing", *IEEE Photonics Technology Letters*, vol. 31 no. 23 1854-1857, 2019.
109. Xu, *et al.*, "Advanced adaptive photonic RF filters with 80 taps based on an integrated optical micro-comb source," *Journal of Lightwave Technology,* vol. 37, no. 4, pp. 1288-1295 (2019).
110. X. Xu, *et al.*, "Photonic RF and microwave integrator with soliton crystal microcombs", *IEEE Transactions on Circuits and Systems II: Express Briefs*, vol. 67, no. 12, pp. 3582-3586, 2020. DOI:10.1109/TCSII.2020.2995682.
111. X. Xu, *et al.*, "High performance RF filters via bandwidth scaling with Kerr micro-combs," *APL Photonics,* vol. 4 (2) 026102. 2019.
112. M. Tan, *et al.*, "Microwave and RF photonic fractional Hilbert transformer based on a 50 GHz Kerr micro-comb", *Journal of Lightwave Technology*, vol. 37, no. 24, pp. 6097 – 6104, 2019.
113. M. Tan, *et al.*, "RF and microwave fractional differentiator based on photonics", *IEEE Transactions on Circuits and Systems: Express Briefs*, vol. 67, no.11, pp. 2767-2771, 2020. DOI:10.1109/TCSII.2020.2965158.
114. M. Tan, *et al.*, "Photonic RF arbitrary waveform generator based on a soliton crystal micro-comb source", Journal of Lightwave Technology, vol. 38, no. 22, pp. 6221-6226 (2020). DOI: 10.1109/JLT.2020.3009655.
115. M. Tan, X. Xu, J. Wu, R. Morandotti, A. Mitchell, and D. J. Moss, "RF and microwave high bandwidth signal processing based on Kerr Micro-combs", Advances in Physics X, VOL. 6, NO. 1, 1838946 (2021). DOI:10.1080/23746149.2020.1838946.
116. X. Xu, et al., "Advanced RF and microwave functions based on an integrated optical frequency comb source," Opt. Express, vol. 26 (3) 2569 (2018).
117. M. Tan, X. Xu, J. Wu, B. Corcoran, A. Boes, T. G. Nguyen, S. T. Chu, B. E. Little, R.Morandotti, A. Lowery, A. Mitchell, and D. J. Moss, ""Highly Versatile Broadband RF Photonic Fractional Hilbert Transformer Based on a Kerr Soliton Crystal Microcomb", Journal of Lightwave Technology vol. 39 (24) 7581-7587 (2021).
118. Wu, J. *et al.* RF Photonics: An Optical Microcombs' Perspective. IEEE Journal of Selected Topics in Quantum Electronics Vol. **24**, 6101020, 1-20 (2018).
119. T. G. Nguyen *et al.*, "Integrated frequency comb source-based Hilbert transformer for wideband microwave photonic phase analysis," *Opt. Express,* vol. 23, no. 17, pp. 22087-22097, Aug. 2015.
120. X. Xu, *et al.*, "Broadband RF channelizer based on an integrated optical frequency Kerr comb source," *Journal of Lightwave Technology,* vol. 36, no. 19, pp. 4519-4526, 2018.
121. X. Xu, *et al.*, "Continuously tunable orthogonally polarized RF optical single sideband generator based on micro-ring resonators," *Journal of Optics,* vol. 20, no. 11, 115701. 2018.
122. X. Xu, *et al.*, "Orthogonally polarized RF optical single sideband generation and dual-channel equalization based on an integrated microring resonator," *Journal of Lightwave Technology,* vol. 36, no. 20, pp. 4808-4818. 2018.
123. X. Xu, *et al.,* "Photonic RF phase-encoded signal generation with a microcomb source", *J. Lightwave Technology*, vol. 38, no. 7, 1722-1727, 2020.

124. X. Xu, *et al.*, Broadband microwave frequency conversion based on an integrated optical micro-comb source", *Journal of Lightwave Technology*, vol. 38 no. 2, pp. 332-338, 2020.
125. M. Tan, *et al.*, "Photonic RF and microwave filters based on 49GHz and 200GHz Kerr microcombs", *Optics Comm.* vol. 465,125563, Feb. 22. 2020.
126. X. Xu, *et al.,* "Broadband photonic RF channelizer with 90 channels based on a soliton crystal microcomb", *Journal of Lightwave Technology*, Vol. 38, no. 18, pp. 5116 – 5121 (2020). doi: 10.1109/JLT.2020.2997699.
127. M. Tan et al, "Orthogonally polarized Photonic Radio Frequency single sideband generation with integrated micro-ring resonators", IOP Journal of Semiconductors, Vol. **42** (4), 041305 (2021). DOI: 10.1088/1674-4926/42/4/041305.
128. Mengxi Tan, X. Xu, J. Wu, T. G. Nguyen, S. T. Chu, B. E. Little, R. Morandotti, A. Mitchell, and David J. Moss, "Photonic Radio Frequency Channelizers based on Kerr Optical Micro-combs", IOP Journal of Semiconductors Vol. **42** (4), 041302 (2021). DOI:10.1088/1674-4926/42/4/041302.
129. B. Corcoran, et al., "Ultra-dense optical data transmission over standard fiber with a single chip source", Nature Communications, vol. 11, Article:2568, 2020.
130. X. Xu et al, "Photonic perceptron based on a Kerr microcomb for scalable high speed optical neural networks", Laser and Photonics Reviews, vol. 14, no. 8, 2000070 (2020). DOI: 10.1002/lpor.202000070.
131. X. Xu, et al., "11 TOPs photonic convolutional accelerator for optical neural networks", Nature **589**, 44-51 (2021).
132. X. Xu et al., "Neuromorphic computing based on wavelength-division multiplexing", **28** IEEE Journal of Selected Topics in Quantum Electronics Vol. 29 Issue: 2, Article 7400112 (2023). DOI:10.1109/JSTQE.2022.3203159.
133. Yang Sun, Jiayang Wu, Mengxi Tan, Xingyuan Xu, Yang Li, Roberto Morandotti, Arnan Mitchell, and David Moss, "Applications of optical micro-combs", Advances in Optics and Photonics **15** (1) 86-175 (2023). DOI:10.1364/AOP.470264.
134. Yunping Bai, Xingyuan Xu,1 Mengxi Tan, Yang Sun, Yang Li, Jiayang Wu, Roberto Morandotti, Arnan Mitchell, Kun Xu, and David J. Moss, "Photonic multiplexing techniques for neuromorphic computing", Nanophotonics **12** (5): 795–817 (2023). DOI:10.1515/nanoph-2022-0485.
135. Chawaphon Prayoonyong, Andreas Boes, Xingyuan Xu, Mengxi Tan, Sai T. Chu, Brent E. Little, Roberto Morandotti, Arnan Mitchell, David J. Moss, and Bill Corcoran, "Frequency comb distillation for optical superchannel transmission", Journal of Lightwave Technology **39** (23) 7383-7392 (2021). DOI: 10.1109/JLT.2021.3116614.
136. Mengxi Tan, Xingyuan Xu, Jiayang Wu, Bill Corcoran, Andreas Boes, Thach G. Nguyen, Sai T. Chu, Brent E. Little, Roberto Morandotti, Arnan Mitchell, and David J. Moss, "Integral order photonic RF signal processors based on a soliton crystal micro-comb source", IOP Journal of Optics **23** (11) 125701 (2021). https://doi.org/10.1088/2040-8986/ac2eab
137. Yang Sun, Jiayang Wu, Yang Li, Xingyuan Xu, Guanghui Ren, Mengxi Tan, Sai Tak Chu, Brent E. Little, Roberto Morandotti, Arnan Mitchell, and David J. Moss, "Optimizing the performance of microcomb based microwave photonic transversal signal processors", *Journal of Lightwave Technology* **41** (23) pp 7223-7237 (2023). DOI: 10.1109/JLT.2023.3314526.
138. Mengxi Tan, Xingyuan Xu, Andreas Boes, Bill Corcoran, Thach G. Nguyen, Sai T. Chu, Brent E. Little, Roberto Morandotti, Jiayang Wu, Arnan Mitchell, and David J. Moss, "Photonic signal processor for real-time video image processing based on a Kerr microcomb", *Communications Engineering* **2** 94 (2023). DOI:10.1038/s44172-023-00135-7
139. Mengxi Tan, Xingyuan Xu, Jiayang Wu, Roberto Morandotti, Arnan Mitchell, and David J. Moss, "Photonic RF and microwave filters based on 49GHz and 200GHz Kerr microcombs", Optics Communications, 465, Article: 125563 (2020). doi:10.1016/j.optcom.2020.125563. doi.org/10.1063/1.5136270.
140. Yang Sun, Jiayang Wu, Yang Li, Mengxi Tan, Xingyuan Xu, Sai Chu, Brent Little, Roberto Morandotti, Arnan Mitchell, and David J. Moss,  "Quantifying the Accuracy of Microcomb-based Photonic RF Transversal Signal Processors", *IEEE Journal of Selected Topics in Quantum Electronics* **29** no. 6, pp. 1-17, Art no. 7500317 (2023). 10.1109/JSTQE.2023.3266276.
141. Kues, M. et al. "Quantum optical microcombs", Nature Photonics **13**, (3) 170-179 (2019). doi:10.1038/s41566-019-0363-0
142. C.Reimer, L. Caspani, M. Clerici, et al., "Integrated frequency comb source of heralded single photons," Optics Express, vol. 22, no. 6, pp. 6535-6546, 2014.
143. C. Reimer, et al., "Cross-polarized photon-pair generation and bi-chromatically pumped optical parametric oscillation on a chip", Nature Communications, vol. 6, Article 8236, 2015.  DOI: 10.1038/ncomms9236.

144. L. Caspani, C. Reimer, M. Kues, et al., "Multifrequency sources of quantum correlated photon pairs on-chip: a path toward integrated Quantum Frequency Combs," Nanophotonics, vol. 5, no. 2, pp. 351-362, 2016.
145. C. Reimer et al., "Generation of multiphoton entangled quantum states by means of integrated frequency combs," Science, vol. 351, no. 6278, pp. 1176-1180, 2016.
146. M. Kues, et al., "On-chip generation of high-dimensional entangled quantum states and their coherent control", Nature, vol. 546, no. 7660, pp. 622-626, 2017.
147. P. Roztocki et al., "Practical system for the generation of pulsed quantum frequency combs," Optics Express, vol. 25, no. 16, pp. 18940-18949, 2017.
148. Y. Zhang, et al., "Induced photon correlations through superposition of two four-wave mixing processes in integrated cavities", Laser and Photonics Reviews, vol. 14, no. 7, pp. 2000128, 2020. DOI: 10.1002/lpor.202000128
149. C. Reimer, et al., "High-dimensional one-way quantum processing implemented on d-level cluster states", Nature Physics, vol. 15, no.2, pp. 148–153, 2019.
150. P.Roztocki et al., "Complex quantum state generation and coherent control based on integrated frequency combs", Journal of Lightwave Technology **37** (2) 338-347 (2019).
151. S. Sciara et al., "Generation and Processing of Complex Photon States with Quantum Frequency Combs", IEEE Photonics Technology Letters **31** (23) 1862-1865 (2019). DOI: 10.1109/LPT.2019.2944564.
152. Stefania Sciara, Piotr Roztocki, Bennet Fisher, Christian Reimer, Luis Romero Cortez, William J. Munro, David J. Moss, Alfonso C. Cino, Lucia Caspani, Michael Kues, J. Azana, and Roberto Morandotti, "Scalable and effective multilevel entangled photon states: A promising tool to boost quantum technologies", Nanophotonics 10 (18), 4447–4465 (2021). DOI:10.1515/nanoph-2021-0510.
153. L. Caspani, C. Reimer, M. Kues, et al., "Multifrequency sources of quantum correlated photon pairs on-chip: a path toward integrated Quantum Frequency Combs," Nanophotonics, vol. 5, no. 2, pp. 351-362, 2016.
154. Hamed Arianfard, Saulius Juodkazis, David J. Moss, and Jiayang Wu, "Sagnac interference in integrated photonics", Applied Physics Reviews vol. **10** (1) 011309 (2023). doi: 10.1063/5.0123236. (2023).
155. Hamed Arianfard, Jiayang Wu, Saulius Juodkazis, and David J. Moss, "Optical analogs of Rabi splitting in integrated waveguide-coupled resonators", **Advanced Physics Research** **2** (2023). DOI: 10.1002/apxr.202200123.
156. Hamed Arianfard, Jiayang Wu, Saulius Juodkazis, David J. Moss, "Spectral shaping based on optical waveguides with advanced Sagnac loop reflectors", Paper No. PW22O-OE201-20, SPIE-Opto, Integrated Optics: Devices, Materials, and Technologies XXVI, SPIE Photonics West, San Francisco CA January 22 - 27 (2022). doi: 10.1117/12.2607902
157. Hamed Arianfard, Jiayang Wu, Saulius Juodkazis, David J. Moss, "Spectral Shaping Based on Integrated Coupled Sagnac Loop Reflectors Formed by a Self-Coupled Wire Waveguide", IEEE Photonics Technology Letters vol. **33** (13) 680-683 (2021). DOI:10.1109/LPT.2021.3088089.
158. Hamed Arianfard, Jiayang Wu, Saulius Juodkazis and David J. Moss, "Three Waveguide Coupled Sagnac Loop Reflectors for Advanced Spectral Engineering", Journal of Lightwave Technology vol. **39** (11) 3478-3487 (2021). DOI: 10.1109/JLT.2021.3066256.
159. Hamed Arianfard, Jiayang Wu, Saulius Juodkazis and David J. Moss, "Advanced Multi-Functional Integrated Photonic Filters based on Coupled Sagnac Loop Reflectors", Journal of Lightwave Technology vol. **39** Issue: 5, pp.1400-1408 (2021). DOI:10.1109/JLT.2020.3037559.
160. Hamed Arianfard, Jiayang Wu, Saulius Juodkazis and David J. Moss, "Advanced multi-functional integrated photonic filters based on coupled Sagnac loop reflectors", Paper 11691-4, PW21O-OE203-44, Silicon Photonics XVI, SPIE Photonics West, San Francisco CA March 6-11 (2021). doi.org/10.1117/12.2584020
161. Jiayang Wu, Tania Moein, Xingyuan Xu, and David J. Moss, "Advanced photonic filters via cascaded Sagnac loop reflector resonators in silicon-on-insulator integrated nanowires", Applied Physics Letters Photonics vol. **3** 046102 (2018). DOI:/10.1063/1.5025833
162. Jiayang Wu, Tania Moein, Xingyuan Xu, Guanghui Ren, Arnan Mitchell, and David J. Moss, "Micro-ring resonator quality factor enhancement via an integrated Fabry-Perot cavity", Applied Physics Letters Photonics vol. **2** 056103 (2017). doi: 10.1063/1.4981392.